\documentclass[12pt]{iopart}
\usepackage[pdftex]{graphicx}
\usepackage{cite}

\usepackage{color}
\usepackage{multirow}

\date{}

\newcommand{\be}{\begin{eqnarray}}
\newcommand{\ee}{\end{eqnarray}}

\begin{document}

\title{Discovery and information-theoretic characterization of transcription factor binding sites that act cooperatively}

\author{Jacob Clifford$^{1,3}$}

\author{Christoph Adami$^{1,2,3\ast}$}
\address{$^{1}$Department of Physics and Astronomy, Michigan State University, East Lansing, MI}
\address{$^{2}$Department of Microbiology and Molecular Genetics, Michigan State University, East Lansing, MI}
\address{$^{3}$BEACON Center for the Study of Evolution in Action, Michigan State University, East Lansing, MI}

\ead{adami@msu.edu}

\begin{abstract}
Transcription factor binding to the surface of DNA regulatory regions is one of the primary causes of regulating gene expression levels.  A probabilistic approach to model protein-DNA interactions at the sequence level is through Position Weight Matrices (PWMs) that estimate the joint probability of a DNA binding site sequence by assuming positional independence within the DNA sequence.  

Here we construct conditional PWMs that depend on the motif signatures in the flanking DNA sequence, by conditioning  known binding site loci  on the presence or absence of additional binding sites in the flanking sequence of each site's locus.  Pooling known sites with similar flanking sequence patterns allows for the estimation of the conditional distribution function
over the binding site sequences.  

We apply our model to the Dorsal transcription factor binding sites active in patterning the Dorsal-Ventral axis of \textit{Drosophila} development.  We find that those binding sites that cooperate with nearby Twist sites on average contain about 0.5 bits of information about the presence of Twist transcription factor binding sites in the flanking sequence.  We also find that Dorsal binding site detectors conditioned on flanking sequence information make better predictions about what is a Dorsal site relative to background DNA than detection without information about flanking sequence features.  

\end{abstract}
\newpage
\section*{Introduction}

The 'particle' abstraction of classical mechanics reduces the many degrees of freedom of an extended material into a single point in space and time~\cite{landaumech}.  A similar abstraction is useful for treating the process of regulation by transcription factor proteins of gene regulatory networks.  In such a model, the entire genome is seen as a one-dimensional lattice where each lattice 'site' is like a type of static particle with a coordinate along the genome, and where the site is a short sequence of DNA, ranging from a single base-pair to a coarse-grained extended sequence of DNA.  Each such site can be defined by its specific logic given by the interactions that are relevant for regulating transcription~\cite{pmid19104053,pmid9169833,pmid17903288}. This logic, encoded in the type of site, is an inheritable trait.  Furthermore, evolution of regulatory sites changes the logic, which is known to cause major transformations on animal body plans~\cite{pmid10892643,pmid12777501}.  Understanding this logic, at a sequence level, has produced state of the art phylogenetic models for classification at the phylum level that allows us to better understand our deepest homologies with the rest of the kingdom.

\subsection*{Position Weight Matrices}
Commonly, estimating the nucleotide frequencies of functional transcription factor binding sites is achieved by aligning experimentally confirmed functional sites of length $s$, and counting the frequency of each nucleotide at each position.  These counts can then be used to infer the distribution of functional binding site sequences. The inferred distribution of functional sequences is called a Position Weight Matrix~\cite{pmid9581503}.  For example, for a length $s$ binding site, the probability that the binding site has the sequence \textit{S} is:
\begin{equation}\label{pwm1}
P(S) = \prod_{ij}^{s,3} P_{ij}^{S_{ij}},
\end{equation}
where the sequence \textit{S} is represented by the matrix of indicator variables $S_{ij} \in \{0,1\}$ (Boolean variables), and $P_{ij}$ is the probability (maximum likelihood estimate from the frequencies) to find base $j$ at position $i$, with $i \in \{1,2,\dots, s\}$ and $j\in \{0,1,2,3\}$, such that each integer represents a letter from the alphabet A,C,G,T.

Information theoretic and classification methods can then be used to relate these probabilities to linear (additive) logarithmic models or a discrimination function.  For example, the energy PWM gives a bioinformatic score \textit{E(S)}, for any sequence \textit{S}.  The energy of the sequence can be decomposed into a sum over each internal position of the sequence:
\begin{equation}\label{epwm}
E(S) =\sum_i^s \sum_j^3 E_{ij} S_{ij},
\end{equation}
where the binding site sequence \textit{S} is again represented by the indicator variable $S_{ij}$ for each position \textit{i} in the sequence and base-pair \textit{j}, which selects the appropriate transcription factor-DNA interaction energies $E_{ij}$.  We define the interaction energies $E_{ij}$ mathematically in \Eref{epwm2} below.

\subsection*{In-vitro Biophysical PWMs}
The energy weight matrix elements used in \Eref{epwm} can be determined for each of the 4$s$ matrix elements using an affinity assay.  This assay is purely based on physical principles, completely blind to notions of ``functional" (meaning adapted) binding sequences.  The key measurement is the relative change in affinity to the transcription factor for all possible single mutation sequences from the highest affinity sequence~\cite{pmid9581503,pmid8080080,hill,pmid20877328,pmid9268651}.  Such an assay assumes that the highest affinity sequence (which we denote as $\rm{S_0}$), is known.   By choosing the highest affinity sequence as the reference DNA-transcription factor interaction, one can then construct the full set of relative affinities for full sequences (all $4^{\rm s}$ affinities).  Just as a key assumption of the PWM model was linearity in sequence, so too in this experiment we must assume that the binding energy is a linear function of the sequence.  This assumption enables each of the three possible DNA mutations from the reference sequence at a particular position within the DNA binding site to be tested independent of the genetic background of the remaining positions within the binding site.

The theoretical justification that the binding energy is a linear function of the sequence is that the binding affinity constant $K(S)$ is equal to the exponential of the binding energy in units of k\textit{T}, where k is Boltzmann's constant and $T$ is temperature.  The free energy, being a state function (i.e., exact differential), then would result in the following displacement reaction: $\log{K(S)} = \log{K(\rm{S_0})} -\Delta G$, where the transcription factor was originally bound to sequence $\rm{S_0}$ (the reference sequence) and then (by any physical process) is displaced and binds to sequence $S$.  If we set the energy scale such that the highest affinity sequence bound to the protein has zero energy, then all other bound complexes have higher energies $G(S)$, hence $\Delta G = G(S)-G(\rm{S_0}) = G(S)$.

Using the physical approach above, one can treat each mutation of a base from the reference sequence (highest affinity sequence) as a perturbation of the reference sequence $\rm{S_0}$.

By expanding the free energy in sequence space, we have
 \be
    G(S) &=  \sum_{ik}^{s,3} \Delta G_{ik} S_{ik} +  \sum_{i,j=1}^s\sum_{k,l}^3 w_{ijkl} S_{ik}S_{jl} + \dots \\
    & \approx \sum_{ik}^{s,3} G_{ik} S_{ik}.
  \ee \label{Gapprox}
The pairwise interaction term, $w_{ijkl}$, is a function of four indices, where indices $i$ and $j$ run over the positions of the sequence $S$, and the indices $k$ and $l$ run over the nucleotide bases.  The indicator variables $S_{ik}$ and $S_{jl}$ select the appropriate pairwise interaction term $w$.  The expansion in sequence space has a total of $2^{\rm s}$ interactions, the final approximation assumes all these are negligible except the first order terms.

\subsection*{Evolutionary PWMs}

Just as a phylogenetic analysis of genes can reveal subsequences that are important for the function or enzymatic activity of the protein, so too can phylogenetic analysis of binding sites reveal subsequences that are important for the binding function (affinity) of the sequence.  Unlike cladistics, where a binding site alignment would only include a monophyletic group (sequences evolved from a common ancestor), and hence be hampered by patterns of conservation that are due to inheritance as opposed to adaptations, here we use a phenetic approach to alignment, based on Berg and von Hippel's phenetic approach~\cite{pmid3612791}, where both convergent sites, paralogs, and orthologs are used in the alignment to reveal conserved patterns in the DNA binding sites that are a consequence of the molecular properties that provide the binding phenotype.

A basic molecular evolution principle initially formulated by Zukerlandyl and Pauli and latter utilized by Dayhoff and refined by Kimura is that neutral DNA accumulates substitutions with a reliable rate, such that \textit{neutral} DNA can be used as a molecular clock.  However, \textit{functional} DNA's mutation rate (what Berg and von Hippel called the ``base-pair choices") are correlated with the functionality of a site~\cite{pmid3612791}.  Hence, functional DNA under purifying selection evolves slower (if at all) than neutral DNA, enabling a comparative analysis of regulatory sequences by screening conserved blocks of sequences, or ``phylogenetic footprints"~\cite{pmid15511292}.

Berg and von Hippel used these assumptions in 1987 to relate the empirical nucleotide counts from an alignment to theoretical binding site sequences under mutation-selection balance~\cite{pmid3612791}.  Theoretically, they assumed a binding site was constrained by the binding affinity necessary for binding (\textit{i.e.}, binding that influences gene expression)~\cite{Berg15081992}.  This constraint allowed them to use Jaynes's principle to derive a theoretical distribution known in physics as the Boltzmann distribution, which they then could equate to the empirical normalized counts from~\Eref{pwm1}.  In this context, Jaynes principle states that the information content of the set of binding site sequences (\textit{i.e.}, binding site sequence data in the form of \Eref{pwm1} and knowledge of the genome-wide frequencies--the prior, or GC content of the genome--should be minimized subject to the binding energy constraint~\cite{hobson}.

For a simple example, consider a binding site of just one base-pair\footnote{Binding sites are frequently about 10 base-pairs long.  A binding site of length one base-pair is not realistic for transcription factors, as most proteins would cover more space than one base-pair (about 1 nanometer).  For an evolutionary argument for why binding sites are about 10bp in length see~\cite{pmid22887818}, and for a diffusion argument see~\cite{pmid21723826}}. The ``Lagrangian" for the constrained minimization problem can be written as (the sum is over the nucleotides that base $B$ can take on)
\begin{equation}\label{lagrangian}
\sum_B{P(B)\log{\frac{P(B)}{P_0(B)}}} -\lambda_0(\sum_B P(B) -1) - \lambda_1(\sum_B P(B)G(B)-\langle G\rangle).
\end{equation}
The first term is the information content of the steady state probabilities $P(B)$ relative to the genome-wide frequencies $P_0(B)$, the prior.  The second term represents the normalization constraint over the probabilities (where the prior is assumed fixed) and the last term is the constraint that the average binding energy be fixed.  Minimizing the Lagrangian leads to the theoretical estimate of the steady state distribution, $P(B)$, which takes the form of a Boltzmann distribution (e.g., see \Eref{bz} for a definition of the Boltzmann distribution).

The equilibrium frequencies, $P_0(B)$, are those expected of neutral DNA (\textit{e g.}, the frequencies estimated from the Jukes Cantor substitution model.  Sites under selection are forced away from equilibrium, and form a steady state distribution $P(B)$.  For a physical example, the relative frequency of a particular base $B$ is like a concentration, which when in thermodynamic equilibrium will be equal to the concentration of this molecule in the background.  Assuming the background can be modeled as chemically random bases (A,C,G,T)~\cite{345660479}, then in thermodynamic equilibrium the base $B$s concentration will equal the background concentration of the respective base.  In a steady state, however, the base frequency is forced to to concentration unequal to the background.  Similarly, in an evolutionary steady state, there is a flux of mutations driving the population of binding sites to the random frequencies, but this flux is balanced predominantly by the flux from the selective pressure.  In the population genetics sense, the steady state frequencies are the result of mutation selection balance. 

 \subsection*{Relation between biophysical PWMs and evolutionary PWMs  }
As a consequence of Berg and von Hippel's hypothesis that the normalized frequencies from an alignment of binding sites could be equated to the theoretical distribution of sequences under mutation-selection balance (the Boltzmann-like distribution)~\cite{pmid3612791}; Berg and von Hippel were able to derive a simple relation between their information theoretic logarithmic score E(S) from~\Eref{epwm}, and the known binding energies $G(S)$ of the binding sites to the transcription factor from \Eref{Gapprox}. Using the standard statistical mechanics relation: $\log{\frac{K(S)}{K(S_0)}}=\log{\frac{P(S)}{P(S_0)}}$, where $K(S)$ is the binding constant, and $P(S)$ is the Boltzmann-like distribution (see \Eref{bz} for details), and observing that $\log{\frac{P(S)}{P(S_0)}}$ can be replaced by the normalized frequencies from the alignment, and defining the information theoretic score from \Eref{epwm} as $E(S)=\log{\frac{P(S)}{P(S_0)}}$\footnote{Here we our conflating our notation for $P(S)$, which in one case is the empirical normalized frequencies from the alignment (which Berg and von Hippel denoted as $f(S)$), while in the other case of statistical mechanics $P(S)$ is a theoretical distribution parameterized by the Lagrange Multipliers (which can be shown to be the thermodyanmic temperature for systems like an ideal gas\cite{Atkins}).  Here we do keep the derived variables $E(S)$ and $G(S)$ separate, in order to clearly see the relation between the bioninformatic score $E(S)$ and the free energy $\Delta G$.}; one then obtains:   

	\begin{equation}\label{bvhke}
	\log{K(S)} = \log{K(S_0)} -\frac{\Delta(E(S))}{\lambda_1},
	  \end{equation}
where $E(S)$ is  estimated from an alignment. (\textit{E(S)} is fully explained in our Methods section, where $\Delta E(S) = E(S)$, and similarly $\Delta G(S) = G(S)$  by choosing $S_0$ to be a reference.) 
The linear relation above is of the same form as the first-order thermodynamic perturbation\cite{hill}:
	\begin{equation}
	  \log{K(S)} \approx \log{K(S_0)} -   \Delta G.
	   \end{equation}
	   This gives us a linear relation between the evolutionary substitution pattern (data from an alignment), $E$, and the free energy, $G$ (in units of k$T$).

\subsection*{Shortcomings of PWMs}

Analyzing typical functional binding site sequences for a particular transcription factor reveals  signs of a conserved pattern of nucleotides at specific positions within the binding site.  However, because the sequences are short, false-positive matches to the pattern are expected to occur frequently in large genomes, too frequently than time available for the protein to find all the sites.  This kinetic search problem was also analyzed by Berg and von Hippel using one- and three-dimensional diffusion models~\cite{pmid7317363}, which has since been reinterpreted several times.  In particular, Sela \textit{et al.} showed that symmetries in DNA sequences flanking functional binding site loci can dramatically affect binding~\cite{pmid21723826}, later verified experimentally~\cite{pmid25313048}.  In the same manner, bioinformatic searches for binding sites using only the conserved patterns in order to discover new binding sites often results in poor predictions on a genomic scale~\cite{pmid14983022}.  

Another limitation of the model is that in development, heterotypic clusters of binding sites (rather than isolated sites) govern gene expression.  Hence, binding site sequence matches to a motif, if occurring in an isolated locus within a genome (i.e., not occurring within a cluster of other binding sites) are incapable of recruiting the complexes necessary for transcription, and hence these isolated loci are unlikely functional.  Hence the functional sequence distribution simply does not contain enough information to make a one-to-one map to the functional loci~\cite{Schneider.Stephens1990}.  Furthermore, in eukaryotes, binding is modulated by the chromatin state of a locus and the cellular state that the genome resides in.  These epigenetic cues and other external variables that influence binding are not usually encoded into the binding site sequences, and gives rise to departures from the linear assumption inherent the PWM model.  

Evolution in development has repeatedly evolved new combinations of binding sites producing new types of logic regulating gene expression~\cite{Gehring1998,Davidson2001,Davidson2006}.  Traditional bioinformatic sequence tools to discover binding sites in developmental systems 
can discover the low resolution segments (500 bp) of regulatory DNA that contain clusters of coevolving binding sites, CRMs, by simply using clusters of motifs~\cite{Brown2008}.  However, determining what sequences within the CRM are functional is difficult. For example: is the spacing between sites functional, is the ordering of sites functional, what about 'half sites' or sites with mismatches, what is the number of mismatches allowable before a sequence is not functional?  Tedious genetic experiments must be conducted in order to discover what sites significantly contribute to gene expression~\cite{Davidson2006}.  

For example, the \textit{in vivo} binding site contribution to gene expression can be understood by comparing the expression of a target gene driven by a wild-type CRM with a knock out of a putative binding site. However, this is complicated for a number of reasons: first, binding site turnover within CRMs leaves remnants of functional sites such as ``half sites" that have partial matches to motifs~\cite{pmid20981027}, second the multiple half sites (that are easier to evolve) may be able to compensate for a strong full site. Therefore, even with a confirmed functional CRM, functional binding site discovery is a daunting task, due to vestigial sites that have fuzzy or poor matches to bioinformatic motifs.
	
\subsection*{Physical Shortcomings of PWMs}
	
  \subsection*{Dependencies within transcription factor bindings sites }
The linear relation in~\Eref{bvhke} becomes nonlinear if there are cooperative interactions between positions within a binding site (or if there are context dependent base-pair dependencies).  For example, cooperativity at the biochemical level tends to cause the linear relationship between the first order Gibbs free energy and the binding constants to become nonlinear as a function of sequence, thereby decreasing the ability of linear models (or first order thermodynamic perturbations) to capture the relationship~\cite{hill,bialek}.  Furthermore, some DNA-protein interactions require specific nucleotides at various positions to jointly occur, such that the additive sum of the interactions of each nucleotide to the protein is not what would be expected under the linear model.  In such cases it becomes important to consider higher-order interactions, such as via dinucleotides or other various joint occurring nucleotides~\cite{pmid20339533,pmid18725950}.

 \subsection*{Dependencies between transcription factor binding sites}
 If the base-pair preferences for a particular transcription factor are contingent on a cooperating factor, then evolution will have filtered the co-occurring sites jointly.  For example, the transcription factor Nf$\kappa$B is known to have a specificity that is dependent on co-occurring binding sites~\cite{pmid15315758}, and similarly the binding sites of the Glucocorticoid Receptor are specific to their context~\cite{pmid19372434}.  The Nf$\kappa$B homolog Dorsal's binding sites have also been shown to encode differences when active in different innate immunity pathways~\cite{busse}, or to signal Dorsal's role as an activator or a repressor~\cite{pmid21890896}.  

\subsection*{Conditional PWMs based on co-occurring factor binding sites}
Here we present a model that incorporates locus-specific information into PWMs that we call ``conditional" PWMs, that improve binding site discovery within CRMs by incorporating flanking information of each binding site locus into the functional binding site sequence distribution.  This is useful for transcription factors that display specialized behavior based on their \textit{cis}-environment.  Our PWM approach accounts for DNA-DNA epistasis (hard-wired cooperativity) that is a function of the DNA spacer between target binding site and a putative cooperating transcription factor's site.  The hypothesis is that base-pair preferences between known cooperating proteins will be a function of the spacer between the known sites (assuming that sites that are separated by large spacers are effectively non-interacting).  If the base-pair preferences change as the spacer changes, then evolution will have filtered the co-occurring sites jointly rather than independently.  As a consequence,  we expect different PWMs for binding sites separated from a putative interacting site as a function of spacer size.  This model is similar to the cooperative nucleotide model in Ref.~\cite{pmid3612791}, but now we effectively have a spacer model between binding sites.  

Furthermore, Berg and von Hippel in Ref.~\cite{pmid3612791} introduce a spacer dependent interaction energy, which similarly addresses that spacing between co-occurring transcription factor binding sites affects the total binding energy between the two separated sites.  However, in their spacer dependent interaction energy, these authors kept the PWM for each binding site a constant, regardless of its interaction with co-occurring binding sites, and only focused on the spacing between the co-occurring binding site.  Our model, in a sense, encodes the spacer dependent interaction energy into the different conditional PWMs constructed for different spacer windows. 
 
\section*{Materials}

\subsection*{Data for known Dorsal binding sites in \textit{D. melanogaster} Dorsal-Ventral network}
The initial development of the fruit fly is partly based on maternally laid morphogens that form a gradient across the blastoderm thereby causing differential target gene expression~\cite{lawr,Hong23122008,pmid15788537}.  The Dorsal-Ventral (DV) network of genes active in the \textit{Drosophila} embryo is largely conserved across the Drosophila genus, furthermore their coarse-grained expression patterns in terms of percent egg length along the DV axis are also largely conserved~\cite{pmid19843594}.  The transcription factor Dorsal regulates the genes responsible for patterning the DV axis of embryogenesis leading to gastrulation~\cite{pmid16271864,pmid16198288,pmid17322397}.  Hence Dorsal transcription factor binding sites within and across {\it Drosophila} species represent a large set of binding sites that are amenable to constructing a PWM.

We collected Dorsal binding sites active in the \textit{Drosophila melanogaster} neuroectoderm region of the DV axis that cooperate with a bHLH (basic helix-loop-helix) dimer with Twist.  These sites are the $\textit{D}_{\beta}$ sites of Table S2 of Crocker et.al.~\cite{pmid20981027}, the Dorsal sites from figure 2 of Crocker \textit{et al.}~\cite{pmid18986212}, as well as the ``specialized" NEE (Neurogenic Ectoderm Enhancers) and NEE-like Dorsal binding sites of Erives \textit{et al.} and Crocker \textit{et al.}~\cite{pmid15026577,pmid18986212}). Those sites are specialized in the sense that they have been shown to evolve slower than flanking Dorsal binding sites in homotypic clusters of Dorsal binding sites in the NEE~\cite{pmid20981027}, and possibly specialized to the cooperative interaction with Twist (which we aim to characterize through information techniques).  

There is ample evidence and a long-standing history in the literature for Dorsal sites cooperating with a bHLH dimer, see~\cite{pmid15128669,pmid1925551,pmid16750631,pmid7774581,pmid8453668,pmid1325394} and references therein. In those cases, the bHLH dimer is likely a Twist:Daughterless heterodimer.  Daughterless is a ubiquitously expressed and obligate partner in tissue-specific bHLH dimers, such as Twist.  The 'specialized' Dorsal data set is labelled as $\mathcal D_{\textbf{DC}\rm{mel}}$, where $\mathcal D$ represents a data set, and the subscript $\textbf{DC}$ means 'Dorsal Cooperative' and mel stands for the species \textit{melanogaster}.

We also collected Dorsal binding sites from the REDFLY footprinting database~\cite{Gallo21102010} for target sites active in embryogenesis.  This data set is labeled as $\mathcal D_{\textbf{DU}\rm{mel}}$, where $DU$ means 'Dorsal Uncooperative'. We did not find Dorsal footprinted sites from REDFLY for the Dorsal target gene \textit{snail} in the CRM of \textit{snail}, hence these Dorsal binding sites were omitted from our data set (our CRM data are described below).   The $\mathcal D_{\textbf{DU}\rm{mel}}$ is a subset of the full REDFLY Dorsal binding sites, where we filtered out any sites that had already been collected in our $\mathcal D_{\textbf{DC}\rm{mel}}$ data set, or sites that were not active in the DV network, or binding site loci that overlapped.

\subsection*{DNA sequence context of binding sites}
Our aim is to characterize the Dorsal sites based on patterns in the loci's flanking sequence.  The regulatory regions (the cis-regulatory modules) of DNA that contain the $D_{\textbf{DC}\rm{mel}}$ and $D_{\textbf{DU}\rm{mel}}$ binding sites consisted of the following \textit{melanogaster} CRMs: \textit{rho, brk, sog, sogS, vn, vnd, twi, zen, dpp, tld}. In that list the CRM is labeled by the gene it targets, and the \textit{sog} gene had its Dorsal binding sites in two distinct CRMs labeled \textit{sog} and \textit{sogS} (where \textit{sogS} is a `Shadow' enhancer).

These CRMs have been collected in a centralized file by Papatsenko et al.~\cite{pmid19651877}.  Additionally, these authors collected known \textit{melanogaster} modules from the literature and using a BLAST approach predicted the remaining 11 Drosophila orthologs of the known \textit{melanogaster} regulatory regions (at that time there we 12 sequenced genomes for \textit{Drosophila}).  The orthologs were not `known' with same certainty as the \textit{melanogaster} data, however we will still classify these as known for our purposes, as conservation of synteny (order of sites) along with each module containing multiple conserved blocks where sequence matches to binding sites reside renders these predictions accurate.  These modules are usually minimal modules that are on average about 300 base pairs in length.

We aligned the 12 orthologs of each CRM, and only extracted the aligned blocks that contained our $D_{\textbf{DC}\rm{mel}}$ and $D_{\textbf{DU}\rm{mel}}$ binding sites, see Supplement section 1.1 for details.  The enlarged set of combined data we call $D_{\textbf{CB}}=D_{\textbf{DC}} \cup D_{\textbf{DU}}$, where the removed subscript mel on DC and DU, denotes that all 12 orthologs of a given binding site sequence are in the data set, and CB stands for combined.  
	

\section*{Methods}

\subsection*{Clustering Dorsal target loci based on co-occurring binding sites}
Given the locations of the Dorsal binding sites within a given CRM (see Supplement section 1.2 for details) and the \textit{predicted} sites of another factor (a putative cooperating factor), we are able to construct a distance matrix where each row `\textit{i}' is a known Dorsal locus (base-pair coordinate), and each column represents a predicted co-occurring factor's locus `\textit{j}' within the CRM.  The matrix elements of the distance matrix are the spacer length (denoted as $d(i,j)$ ) in base-pairs between any row \textit{i} (Dorsal binding site locus) and column \textit{j} (co-occurring binding site locus), a difference of the coordinates $z$ of the loci:
\begin{equation}\label{cluster}
d(i,j) = z^i - z^j -w^i,
\end{equation}
where we assume that the \textit{i}th Dorsal site appears upstream from the \textit{j}th co-occurring site, and that both sites are annotated as on the positive strand of the CRM, where $w^i$ is the width (length) of the \textit{i}th site, and $z^i$ and $z^j$ are the CRM coordinates of site $i$ and $j$ respectively.  Here we define the spacer as the base-pair distance of neutral DNA between two binding sites (hence the internal positions within either site are not counted as part of the spacer).  For cases where the Twist and Dorsal site overlap, the spacer is valued at 0 bp regardless of the amount of overlap.  For cases that a CRM did not contain a predicted co-occurring site, we set the spacer to a maximum value such that the corresponding Dorsal site for the spacer was guaranteed to be classified as "Uncooperative".

\subsection*{Classifying binding sites based on spacer window}

We define a partitioning of the flanking sequence of any given Dorsal locus, hence we use the reference frame of the Dorsal locus with both upstream and downstream sequence.  We partition the upstream flanking sequence by the minimum distance $d_{\rm min}$ and a maximum distance $d_{\rm max}$ away from the locus using \Eref{cluster}.  Similarly, we define a symmetric partition of the downstream flanking sequence by the minimum distance -$d_{\rm min}$ and a maximum distance -$d_{\rm max}$ away from the locus.  We then define a coarse-grained binning of all the flanking sequence into just two bins, where a `spacer window' represents the bin that contains the interval $[d_{\rm min}, d_{\rm max}] \cup [-d_{\rm min}, -d_{\rm max}]$, and the other bin contains all the rest of the flanking sequence.  Once the bin borders have been defined by the spacer window, we then define a Boolean class variable $C$, which classifies each Dorsal locus as $C=1$ if the co-occurring binding site of interest is present \textit{in} the spacer window, and $C=0$ if the co-occurring binding site sequence of interest is absent \textit{in} the spacer window.  Hence, the class variable is entirely based on the patterns that occur within the spacer window, as the class value of each class is determined solely on co-occurring sites in the spacer window.  Using \Eref{cluster} we classify the Dorsal loci that fall within a defined window.  Once each Dorsal binding site's locus is assigned a class, we then can align the loci of a class and estimate the conditional PWM. 

\subsection*{Energy estimation of a base}
The theoretical steady-state Boltzmann-like distribution is the solution to minimizing the Lagrangian with respect to $P(B)$ in \Eref{lagrangian}.  The Boltzmann-like distribution in units of the second Lagrange multiplier is:
\begin{equation}\label{bz}
P(B)=\frac{P_0(B)\exp{-E(B)}}{Z},
\end{equation}
where the normalization $Z$ is related to the Lagrange multiplier $\lambda_0$, and we have assumed calibration of the energy $E(B)$ by estimating the shift and scaling factors from \Eref{bvhke}.  
Assuming our frequencies from \Eref{pwm1} is governed by the Boltzmann-like distribution, 
we then can construct an energy PWM by inverting the distribution, arbitrarily choosing the consensus base $\rm{B_0}$ to be the zero of the interaction energy between transcription factor and bases.  The consensus base is the base at a position with the most counts from the alignment, hence this choice of reference leads to all other bases contributing a higher energy (or zero for degenerate cases).  We then can calculate the interaction energy of the remaining bases \textit{B} as:
\begin{equation}\label{epwm2}
E(B) \approx -\log{\frac{P(B_0)}{P(B)}} =- \log{ \frac{n_{B_0}+\beta}{n_B+\beta} }.
\end{equation}
Here we have made the approximation that the degeneracy factors $P_0(B) = g(B)/L$ are negligible (this is the prior or background DNA frequency), where $g(B)$ is the multiplicity or number of times that base $B$ occurs in a genome of length $L$~\cite{pmid2082934}, $n_{B_0}$ are the counts of the reference base $B_0$ and similarly $n_B$ are the counts of base $B$ from the contingency table estimated from the alignment of $n$ known sites, and $\beta$ is a pseudocount $\beta>0$.  The joint energy of a given base $B$ with co-occurring flanking sequence \textit{S'} (that may or may not contain a co-occurring binding site of another factor) is defined as $E(B,S') = E(B) +E(S') + w(B,S')$.  By setting a spacer threshold (spacer window) and an energy threshold on the potential cooperating factor we effectively create a Bernoulli variable for the flanking sequence, such that \textit{S'} is aggregated into the class variable \textit{C}.  Hence, we have $E(B,C) = E(B) +E(C) + w(B,C)$, where \textit{w(B,C)} is an interaction energy that is shared between the systems \textit{B} and \textit{C}.  Once we have determined what class a Dorsal locus belongs to, we are then uninterested in the energy of the co-occurring site in sequence \textit{S}'.  Hence we define a conditional energy that is the standard PWM energy from \Eref{epwm} for a particular position and base plus the interaction term:
\begin{equation}
E(B|C) =E(B)+ w(B,C).
\end{equation}
The interaction term shifts the standard energy of a sequence if $P(B|C) \neq P(B)$.  We define our context \textit{C} for Dorsal sites \textit{B} based on proximity to Twist (the spacer window),  thereby placing a class tag \textit{C}, on each of Dorsal binding site bases \textit{B}.   We calculate the shift \textit{w} as:
\begin{equation}
w(B,C) =- \log{ \frac{P(B,C)}{P(B)P(C)}}=-\log{\frac{P(B|C)}{P(B)}}.
\end{equation}
The shift \textit{w} is simply the Kullback-Leibler divergence of the conditional distribution $P(B|C)$ and the marginal distribution $P(B)$.  

\subsection*{Energy estimation of a sequence of bases}
We now extend the model from a single site to a binding site sequence.  The total shift for a particular binding site sequence $S$ and its flanking sequence is: $w(S,S') = E(S,S') -E(S) -E(S')$, where the shift is calculated as:
\begin{equation}
w(S,S')=  -\log{ \frac{P(S,S')}{P(S)P(S')}}\;.
\end{equation}
The sequence \textit{S} is the Dorsal binding site at a particular locus, and is a sequence of bases \textit{B}, while the sequence \textit{S'} is effectively a Bernoulli variable \textit{C}, which means the flanking sequence \textit{S}' of the Dorsal site either has a Twist site (in which case \textit{C}=proximal) or not, in which case \textit{C}=distal.  Hence, \textit{w(S,S')}= \textit{w(S,C)}, which we define as:
\begin{equation}\label{mibase}
w(S,C)= \sum_i^s w(B_i,C)
\end{equation}
where we have defined \textit{S} as the sequence $\{B_i\}$, where $ i \in \{1,2,3\dots s \}$, and $s$ is the length of the binding site sequence \textit{S}.  \Eref{mibase} uses a standard PWM to calculate \textit{P(S)} (as opposed to using the marginal of \textit{S} over \textit{C}), because the marginal distribution is a ``mixture model" that cannot be factorized into a product of base specific probability factors~\cite{barasch}.
  Computationally, for energy PWMs there is a matrix \textit{w} for each class value of \textit{C}.  By adding the \textit{w} matrix to the energy matrix \textbf{E} (matrix elements defined by \Eref{epwm2}) we obtain a {\em conditional} energy.  We define a {\em conditional detector}, or a {\em conditional energy PWM}, which we use for bioinformatic predictions and annotations of binding site sequences.  The detector trained from sequences of class $C$ then will score each sequence $S$ as:
\begin{equation}\label{cepwm}
  E(S|C) = E(S) + w(S,C)\;.
  \end{equation}
  Here \textit{E(S)} is from \Eref{epwm}, where the matrix elements $E_{ij}$ are equal to $E(B(j)_i)$ from \Eref{epwm2}.  The function \textit{B(j)} is a map between base \textit{B}'s alphabet A,C,G,T and the values of the matrix index \textit{j}: 0,1,2,3; where we define the 0 index to be the consensus base and therefore reference energy level (the ground state).  The matrix index \textit{i} denotes the position of the base, which we previously denoted as $B_i$ in \Eref{mibase}, where it was clear which particular base \textit{B} resided at position \textit{i} of sequence \textit{S}.  Hence the conditional energy is $E(B|C)=-\log{\frac{P(B_0)}{P(B|C)}}$, where $B_0$ is the consensus base of the position independent PWM from \Eref{bz}.

\section*{Model detectors}

 We define two types of Dorsal binding site sequence models (``detectors") that we use for detection and classification.  The first detector is conditioned on flanking sequence motifs, and hence potentially can better resolve functional loci.  The second detector is simply a standard (unconditional) PWM model, which we use as a baseline for model comparison.  
  
First we define the detector that incorporates flanking sequence information.  As we will see,the detector acts like a logic-like gate that we call the ``OR gate", due to its similarity with a standard digital OR gate used in electronics.  The input to the gate is a k-mer, and the output is a decision on whether the k-mer is a Dorsal binding site or just random background DNA.  The detector's decision is based on the conditional energy PWM scores from \Eref{cepwm} described above, that is, its output depends on the output of two distinct ``subdetectors", which we call DC (``Dorsal Cooperative") and DU (``Dorsal Uncooperative").  The DC component of the OR gate scores all incoming k-mers based on the conditional energy for a sequence with class type `proximal', while the DU component scores all incoming k-mers based on the conditional energy for the class type `distal'. The ``OR gate" detector fires (that is, predicts a Dorsal site), if either the DC or the DU detector (or both) fire. 
 In general, any energy PWM model (and hence our conditional energy PWMs) can be used as a linear classifier for binding site sequences.  This classification is based on the following linear equation for any given k-mer:
  \begin{equation}
   y(\textbf{S})=E_c - \textbf{E} \bullet \textbf{S},
   \end{equation}
  Here \textbf{E} and \textbf{S} are vectors from a $4k$ dimensional real vector space, where we elevated the matrix of indicator variables from \Eref{epwm2} to be a bona fide vector. $E_c$ acts as bias that shifts the hyperplane that separates putative functional sites from non-functional sites.  The Euclidean dot product between the two vectors, $\textbf{E} \bullet \textbf{S}$, is defined as the sum over element-wise multiplications, where the energy \textbf{E} is now another vector in the space that projects each k-mer \textbf{S} onto a line of length \textit{E(S)} (i.e., \Eref{epwm}).  The so-called bias or energy threshold is a positive real number ($E_c$), and represents a partitioning of the line defined by \textit{y} into positive and negative real numbers.  Here all k-mers with a positive value of \textit{y} have energy less than $E_c$, and are classified as a binding site.  All k-mers with a negative value of \textit{y} have energy greater than $E_c$, and are classified as random DNA sequence.

The OR Gate detector is partially defined once the flanking sequence feature (the co-occurring binding site motif) and the spacer window have been set (or optimized), as described in the Methods section above.  These settings allow us to estimate the conditional probabilities.  
Hence, using only Dorsal binding site sequences from the data set $D_{\textbf{CB}}$ we are able to train and define an OR Gate that is not mixed with binding sites based on purely bioinformatic matches. 
The second model is the standard PWM, which we call the \textit{CB} detector, where $CB$ stands for the ``combined" set (meaning the conditional and unconditional data sets combined), which we denote by $\mathcal D_{\textbf{CB}}$ ).  The CB model assigns an energy score $E(S)$ to each sequence $S$ as in \Eref{epwm}, which has a corresponding probability $P(S)$ as in \Eref{pwm1}.

\section*{Results}
	  \subsection*{Optimal spacer window for the OR Gate detector}
In order to calibrate our conditional detectors we must define an optimal interval of the spacer window by calculating the mutual information between the known Dorsal binding sites and the potential cooperator's binding site (\textit{eg.}, co-occurring Twist sites).  The spacer window that leads to the maximum mutual information determines an optimal clustering of the Dorsal loci into two classes, which we then can use to build the OR gate.

We predict 5'-CAYATG loci (putative Twist sites) within the CRMs by scoring the CRMs with an energy PWM and threshold that corresponds to exact matches of the Twist motif 5'-CAYATG, which we found to have the highest mutual information with Dorsal binding site sequences.  In the Supplemental Results section titled 'Additional Experiment Supplement' we show a similar analysis with the alternative Twist motif 5'-CACATG, and some results for the motif's restricted form 5'-CACATGT.

Upon construction of the spacer distance matrix we are able to classify all annotated Dorsal sites as `Cooperative' or `Uncooperative', based on whether any of the spacers for a given Dorsal locus was within the bin border defined by $d_{\rm min}$ and $d_{\rm max}$.  For example, a CRM annotated with one Dorsal site and three Twist sites will have three spacers.  If any of those spacers are within the spacer window, then the Dorsal site is classified as `Cooperative'.  We define the spacer window as a 30 base-pair closed interval, which starts at [0,30]bp relative to each Dorsal coordinate within the CRM (not counting the body of the binding site as a part of the spacer).

All known Dorsal loci of a given class are then aligned (see Supplement section 1.6 for details) to construct the conditional Dorsal binding site sequence distribution (conditional PWM).  Given the class labels on the Dorsal sites, we are able to estimate the probability of a given class as simply the fraction of Dorsal sites that belong to each class \textit{C}.  With these distributions we are then able to calculate the mutual information, \textit{I(S;C)} between the Dorsal site sequence variable \textit{S} and the class \textit{C} as
\begin{equation}\label{mi}
I(S;C)= \sum_S \sum_C P(S|C) P(C) \log{\frac{P(S|C)}{P(S)} }
\end{equation}
where $P(S|C)$ is the conditional PWM, and $P(S)=\sum_C\prod_i P(C)P(S_i|C)$ is the marginalized distribution of sequence over class labels \textit{C} (note this is not the same as the CB detector's probability).  As stated above, the initial $d_{\rm min}$ was set at zero and $d_{\rm max}$ at 30bp, and then both parameters are incremented by 30bp to shift the window to a new position.  For each shift of the spacer window we classify all Dorsal loci, align each class to a length 9 motif, and then calculate the mutual information. The result is shown in \Tref{tabless} and implies that the information between sequence and class label is highest if the spacer is between 0 and 30 bps, as expected for binding sites that interact via molecular interactions.  Furthermore we appended one nucleotide of flanking sequence on each binding site sequence to see if we were missing flanking parts of the conditional binding sites.

\begin{figure}
\includegraphics[width=6in]{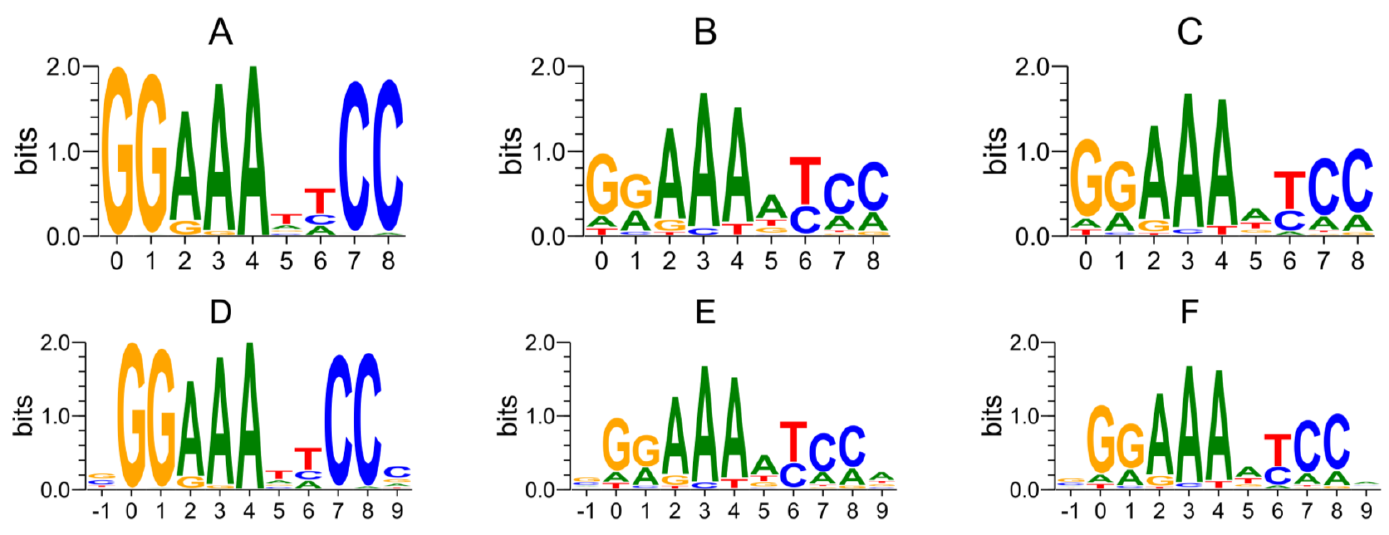}  
\caption{Logos generated for known Dorsal sites (the $D_{CB}$ data) tested for adjacency to 5'-CAYATG used as the cooperative class if in the [0,30]bp distance.  Logo A corresponds to the cooperative class, and displays the known 5'-AAATT core, with total information content 13.5 bits. 
 Logo D is the exact same logo as A but with a single base-pair of flanking sequence at the start and end of the site (hence, this logo starts at position -1).  Position 9 of this logo shows about two decibits of information relative to the background sequence in the nucleotide base `C' (2 out of 10 functional DC sites have a `C' at this position).  Logo B is the `uncooperative' class for the [0,30]bp window, which we calculated to have 9.1 bits information relative to the background (uniform distribution of bases), and logo E has the added flanking sites to the `uncooperative' class.  Logo C is the CB motif with 9.6 bits of information relative to the background, which looks similar to the 'uncooperative' class at position 6 due to there being many more sites that prefer A to a T at this position amongst all the Dorsal sites in the network.  Logo F is the CB motif with the flanking sequence appended.}
\label{logofig}
\end{figure}

\begin{table}[htbp]
\centering
  \begin{tabular}{|c|c|c|c|}
\hline
spacer   &  [0,30]bp & (31,60]bp &(61,90]bp  \\ \hline
 Mutual Information, \Eref{mi} &  0.49 & 0.29 & 0.04  \\
\hline
\end{tabular}
\caption{Mutual Information between functional Dorsal binding site sequences and putative Twist sites that match 5'-CAYATG using a sliding spacer window scheme.}\label{tabless}
\end{table}
We show the conditional Dorsal binding site sequence logos for functional binding sites generated for this first spacer window in \Fref{logofig}. The information content of each position of the binding site corresponds to the height of the logo, where we used a symmetric hyperparameter value of $\beta=0.1$ as discussed in the Supplement section 1.9 and section 1.17. 

\subsection*{The conditional and unconditional PWMs are significantly different}
 
Here we test the optimal DC and DU detector's training data energy scores to see if the median energy of DC is significantly different than the median energy of DU.  The optimal detectors were based on the 5'-CAYAGT Twist motif and the [0,30]bp window.  The rank sum test rejected the null hypothesis that the median energies are equal with  $p=10^{-26}$.  The median energy of the DC PWM was 0.27, while the median energy of the DU PWM was 2.7.

\begin{figure}
  \includegraphics[]{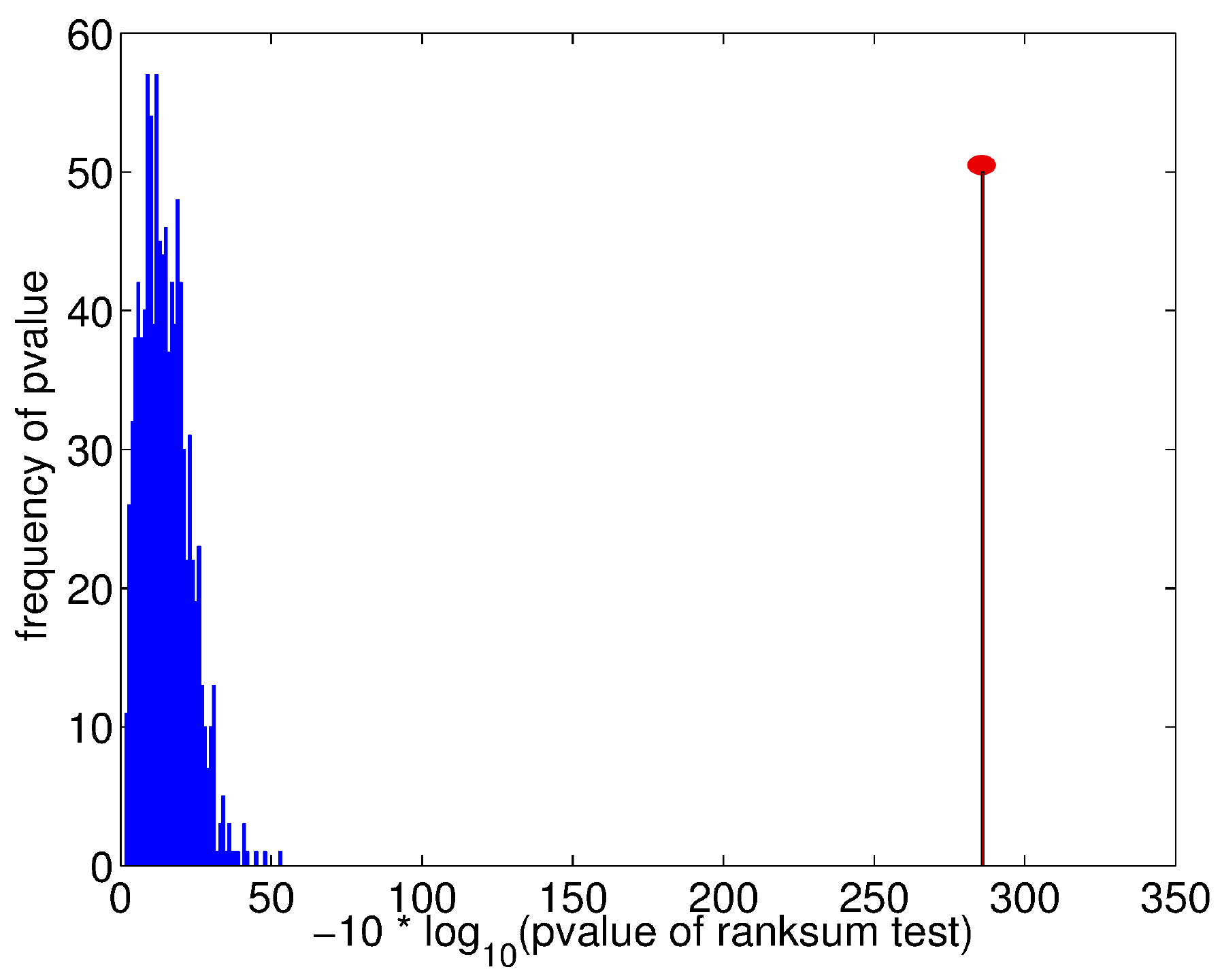}\\
  \caption {Histogram of p-values of a rank sum test of random partitions of the combined data set $\mathcal D_{\textbf{CB}}$.  The binning is in units  $-10\times \log_{10} $ of the p-value, rounded to the nearest integer. The p-value of the rank sum test between DC and DU energy data sets based on their energy PWMs was 260 in log base ten units (scaled by 10), which is indicated by the red bar of arbitrary height.   
   }\label{b}
\end{figure}

It is possible that any random partitioning of a set of binding sites that are used to build detectors using our technique would produce p-values consistent with significance.  We used our original data set of Dorsal sites $\mathcal D_{\textbf{CB}}$ to construct a sampling distribution of p-values for the rank sum test 
 To calibrate the p-value we created a sampling distribution of the p-value from 1000 repetitions, where at each repetition the combined data $\mathcal D_{\textbf{CB}}$ were randomly partitioned into two data sets.  PWMs were constructed for each partition.  The energy of each sequence within a partition was calculated as $E(S)+w(S,P)$, where $P$ is the partition, $S$ is a sequence in the partition, and $E(S)$ is the CB energy.  We then determined the corresponding rank sum p-value between the random sets. We found that the p-value of the rank sum test between the DC and DU model fell  well beyond the right tail of the random sampling distribution (shown in \Fref{b}), indicating that the median energies of DC data set and the DU data set are significantly different from {\em any} random partitioning of the combined data set. More details are in the Supplement section 1.11. 

\section*{\textbf{Performance of optimal classifiers (detectors) }}

All detectors were built from length 9 alignments (see Supplement section 1.6 for details of the alignment procedure).
The OR gate is based on the DC detector built from the data set $\mathcal D_{DC}$, which contains Dorsal loci from  $\mathcal D_{\textbf{CB}}$ that were tagged with class labels from the optimal spacer window of [0,30]bp with the 5'-CAYATG motif, and similarly, the DU detector is built from the data set $\mathcal D_{DU}$, which contains the remaining Dorsal loci from  $\mathcal D_{\textbf{CB}}$ that did not have the Twist sites in the spacer window.  .  
The unbolded subscripts DC and DU on the data sets denote that these sets of Dorsal sites were based on our clustering scheme (not based on literature annotation). 

We now present three experiments that tests the performance of our OR gate detector and the conditional detectors using the CB PWM as a benchmark.

\subsection*{The DC detector predicts sites proximal to 5'-CAYATG with better odds than the DU detector.}
We expect that DC should predict Dorsal binding site sequences that are adjacent to Twist more precisely than DU (since we showed earlier that the Dorsal site sequences contain information about adjacency to Twist).  In \Tref{tabless2} we collected all the hits (all the positives) of the detectors.  We test whether the DC conditional energy PWM is actually \textit{predicting} Dorsal sites within the CRMs that have the correct flanking sequence feature (presence or absence of Twist motif) with better odds than the DU detector.
The odds of DC for predicting binding site sequences that belong to the proximal class was $\frac{61}{39}=1.6$.  The odds of DU for predicting sequences of the proximal class is $\frac{280}{345}=0.81$, hence the odds ratio is 2.0.  The one-sided p-value for this table's log odds ratio test is $p=0.001$ for the chances of seeing a DC detector with better odds relative to DU at predicting correct flanking sequence features.  Increasing the energy cutoff $E_c$ increases the total counts of the table, and we obtain similarly significant tables up until about $E_c=5$.

\begin{table}[h!]
\centering
  \begin{tabular}{ r|c|c| }
\multicolumn{1}{r}{}
 &  \multicolumn{1}{c}{proximal}
 & \multicolumn{1}{c}{distal} \\
\cline{2-3}
$DC$ & 61 & 39 \\
\cline{2-3}
$DU$ &  280 & 345 \\
\cline{2-3}
\end{tabular}
\caption{Contingency table with the conditional detectors DC and DU represented along the rows and the class type distal and proximal represented along the columns. Each table element represents the number of sites predicted from each detector of each class type based on Twist sites (5'-CAYATG) and a CB energy cutoff $E(S)=E_c=2.1$.  
 }\label{tabless2}
\end{table}

\subsection*{Both OR gate and CB detectors show high sensitivity with known sites as positives and CRM sequences as negatives }
In order to test the sensitivity and the specificity of the detectors we used the Receiver Operator Characteristics (ROC), which displays the tradeoff between optimizing predictive performance for `positives', while also optimizing for not detecting known `negatives'.  The True Positive Rate ($TPR$) is defined as $TPR=\frac{TP}{(TP+FN)}$, where the denominator is the total counts of True Positives ($TP$)and False Negatives ($FN$)).  The False Positive Rate ($FPR$) is defined as $FPR=\frac{FP}{(FP+TN)}$, where the denominator is the total counts of True Negatives ($TN$) and False Positives ($FP$).

We use the data set $\mathcal D_{\textbf{CB}}$ as our training set of `positives' ($TP+FN$) for both the CB detector and the OR gate.  The `negative' data set ($TN+FP$) is the set of all CRMs that contained a known binding site (i.e., the CRMs associated with $\mathcal D_{\textbf{CB}}$), where the bona fide sites (the functionally confirmed sites) are masked out.  Furthermore, within the CRMs we also mask out overlapping predicted binding sites based on the algorithm in the Supplement section 1.3, hence the negative data (the CRMs with known sites masked and overlapping hits masked) is at least nine fold smaller than the concatenated length of the CRMs due the binding sites being nine base pairs in length. 

For a given energy threshold, $E_c=E(S)$, set by the CB energy PWM for both the OR gate and the CB detector, each detector `scans' the CRM using a sliding window approach, where each `hit' of the detector is classified as a $TP$ if the hit overlaps a known binding site locus in $D_{\textbf{CB}}$, and as a $FP$ if the detector `misfired' in the background of the CRM.  Similarly, known sites (loci) from $D_{\textbf{CB}}$ that were not called hits by the detector are classified as $FN$, while $TN$ are the k-mers from the CRM background sequence that the detector did not call a hit.

\begin{figure}
\includegraphics[width=6in]{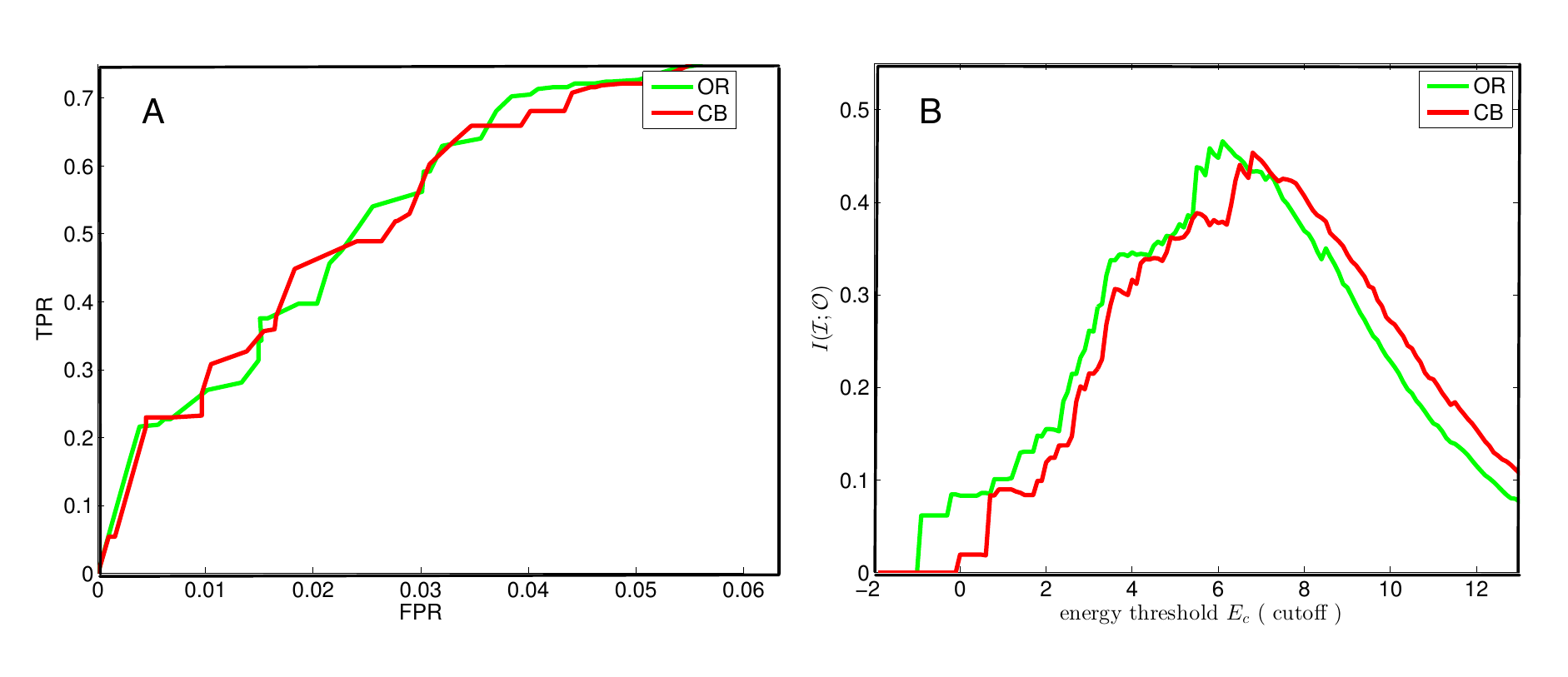}
\caption{ROC and Information. (A) False positive rate ($FPR$) vs. True Positive Rate ($TPR$) when varying the energy cutoff $E_c$.  (B) shows the mutual information $I(\mathcal I;\mathcal O)$ Eq.~(\ref{inout}) between the input and output of the detectors as a function of the cutoff energy. }\label{rocfig}
\end{figure}

The ROC of the OR gate (shown in \Fref{rocfig}A) tends to perform better than the CB detector at low energies up until the energy reaches about $E(S) < 8$ (the last point ($FPR,TPR$) displayed in the figure), after which the CB detector tends to do better.  The OR gate in the region of ROC space displayed shows better performance than the traditional CB detector (This is clearer quantitatively, where we found the OR gate had a higher area under the curve (AUC) integrated from the minimum energy to CB's energy cutoff of $E(S) < 8$ (which is the last point displayed in ROC space)).  The OR gate and CB detector both perform well for strong sites (low energy sites), which is indicated by their good $TPR$ (almost 80\% before a noticeable fraction of negatives start to be detected as positive.

\subsection*{The OR gate performs better than CB at predicting known sites at lower energies}
 Another metric of performance of the classifiers is the mutual information between the type of k-mer (Dorsal or not Dorsal) and the classification by the detector. For example, if the input is not a Dorsal binding site, the detector should stay silent, while it should fire if it is a Dorsal site (either adjacent to Twist or not). We can write this mutual information as
  \begin{equation}\label{miio}
  I(\mathcal I;\mathcal O)=H(\mathcal I)-H(\mathcal I| \mathcal O),
  \end{equation}
 where $\mathcal I$ is the binary random variable holding the true identity of the `Input' k-mer received by the detector, while the 'Output' variable $\mathcal O$ is the binary variable given by the detector's decision. The entropy $H(\mathcal I)$ is in principle given by the relative likelihood to find Dorsal binding sites within the ensemble of CRMs, which is of course heavily biased towards negatives (non-Dorsal sites). However, this Bayesian prior is not available to the transcription factor, in other words, for each decision to bind, the factor has its own Bayesian prior $p$, which we will set to $p=1/2$ (maximum entropy Bayesian prior) below.

The conditional entropy $H(\mathcal I| \mathcal O)=-\sum_{i,o=0}^1p(i)p(i|o)\log p(i|o)$
quantifies the remaining uncertainty about the identity of the k-mer given the decision of the detector, and can be calculated using the false positive and true positive rates introduced earlier. In particular, the conditional probability $p(i|0)$ is obtained as
 \be
    p(1|1)=p(\mathcal I=1|\mathcal O=1) &=  TPR \\ 
   p(1|0)=p(\mathcal I=1|\mathcal O=0) &= 1-TPR\\
   	p(0|1)=p(\mathcal I=0|\mathcal O=1) &= FPR \\
   p(0|0)=p(\mathcal I=0|\mathcal O=0) &= 1-FPR \;,
  \ee 
while  $p(i)$ is the Bayesian prior (density of Dorsals/non-Dorsals in the CRM). Using an arbitrary prior $p$,   
we can rewrite the mutual information from \Eref{miio} as:
\begin{equation}
I(\mathcal I;\mathcal O)=H[p] - pH[TPR] -(1-p)H[FPR]\;,
\end{equation}\label{inout}
where $H[*]$ is the usual binary entropy function of a Bernoulli distribution characterized by *, so for example
 \begin{equation}
 H[TPR] = -\frac{FN}{TP+FN} \log{\frac{FN}{TP+FN}} -\frac{TP}{TP+FN} \log{\frac{TP}{TP+FN}}\;,
 \end{equation}
with a similar expression for $H[FPR]$.
We show the mutual information $I(\mathcal I;\mathcal O)$ in \Fref{rocfig}B using the maximum entropy Bayesian prior $p=1/2$.  Compared to the information the CB detector has about Dorsal sites, the OR gate's information is shifted to lower energies, implying that at fixed energy cutoff it knows Dorsal sites better than CB. 

\subsection* {DC conditional detector is able to predict that Twist is nearby} 
The conditional detectors are expected to make predictions not only about what is a Dorsal site relative to the background, but also whether Dorsal is in the vicinity of Twist.  By partitioning all the known sites into the two class types (e.g., `distal' and `proximal') as determined from the spacer window of [0,30]bp and Twist motif 5'-CAYATG, we can test how well each detector can resolve the class type of a Dorsal site (Dorsal with Twist or without).  

\begin{figure}
  \includegraphics[]{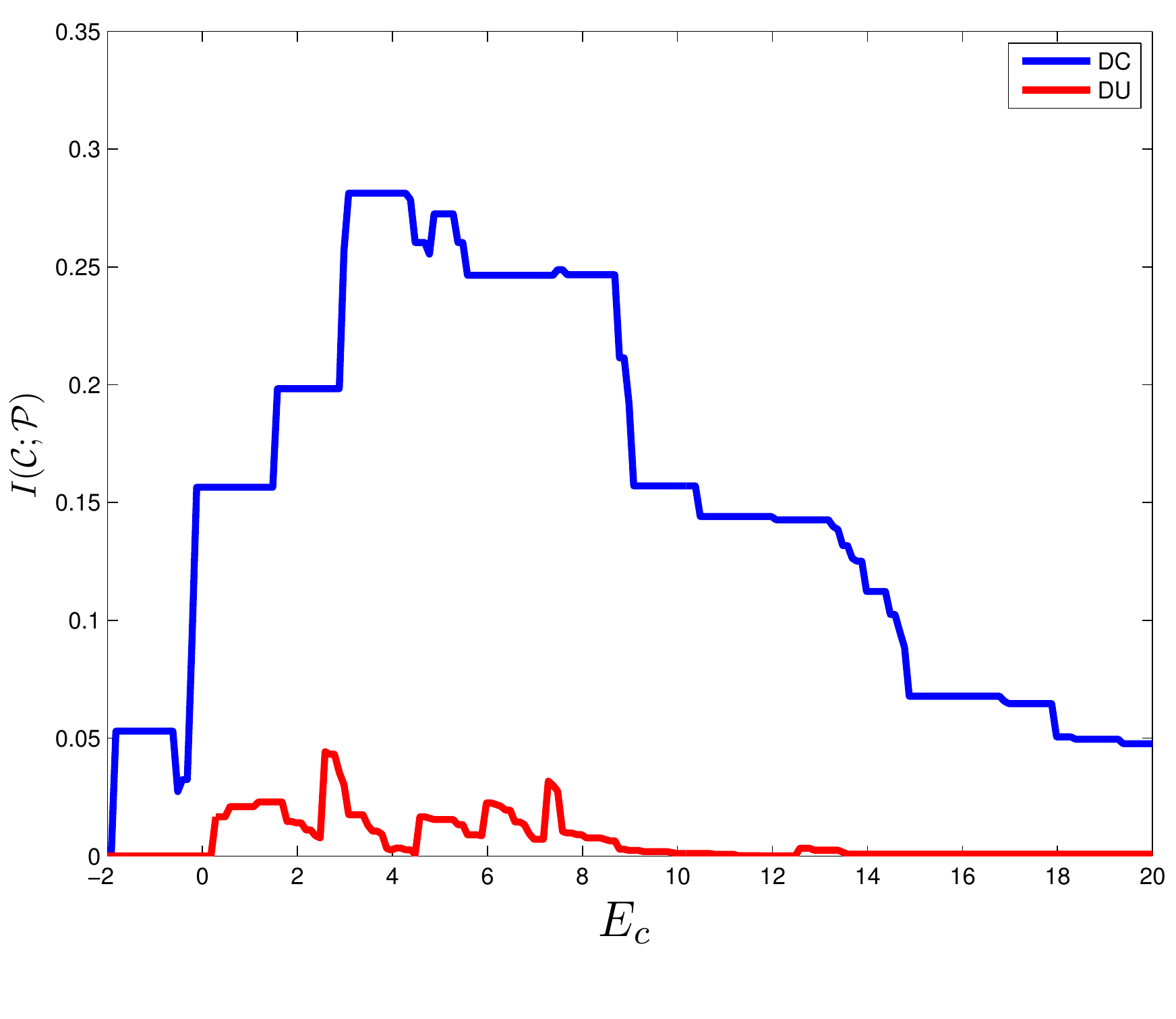}\\
  \caption{Mutual information $I(\mathcal C;\mathcal P)$ between the actual classes $\mathcal C$ and the predicted classes $\mathcal P$ for Detectors DC and DU as a function of the threshold energy $E_c$ that is defined by each detector's conditional energy \Eref{cepwm}.
    \label{MiSP}}
\end{figure}

For a given energy threshold we scanned the combined data set $D_{\textbf{CB}}$ with the DC as well as the DU detector, and asked how much the detector knows about the class variable $\mathcal C$ (further details of this experiment are in Supplement section 1.13).  We show this mutual information $I(\mathcal C;\mathcal P)$ in \Fref{MiSP} , where $\mathcal P$ is the binary random variable encoding the detector's decision about the context. 
We see that the DC detector has up to 0.3 bits of information about the proximity of Twist in any particular Dorsal site, while the DU detector has virtually no information about this variable. 

\section*{Discussion}

\subsection*{DC and DU Information logos and previous evidence}
The binding site sequence logos display the information content of our binding site data relative to a uniform distribution.  By inspection of the DC logo the consensus sequence (highest information scoring sequence) is partially consistent with Table S2 of Crocker et al.~\cite{pmid20981027}.  The 5'-AAATT core is reproduced as our DC consensus sequence, while the flanking sequence for the length 11 binding sites are not enriched with G at the start of the site and a C at the end of the site.  Similarly we can see that our DU also conforms roughly to A-tract Dorsal binding sites, which are Dorsal binding sites that have four or more contiguous Adenines.  Mrinal pointed out that A-tract binding sites have certain physical chemical properties not seen in 5'-AAATT core Dorsal sites~\cite{pmid21890896}, namely that A-tract Dorsal binding sites encode a mechanism (like an extra hydrogen bond between the protein and DNA) for Dorsal to switch roles from an activator of gene expression to a repressor of expression based on the binding site Dorsal was occupying.  Of course, as mentioned by Mrinal, these sites are still context dependent, namely the context of a site may override any preference a binding site sequence has for causing activator or repressor roles\cite{pmid1582412}.  Inspection of our DU detector's data set shows that it is more than 50\% enriched with Dorsal sites that are known to be from repression \textit{cis}-regulatory elements (\textit{zen}, \textit{tld}, \textit{dpp}), hence the DU logo with a 5'-AAAAT core is not surprising.

Our known binding sites, to a degree, come with the class labels already attached.  The $\mathcal D_{\textbf{DC}\textit{mel}}$ data is the known Dorsal binding site data set based on the definition of $\textit{D}_{\beta}$ or 'specialized' sites, or NEE-like Dorsal binding sites (neuroectoderm Dorsal sites that were linked to Twist sites, but were not linked to the canonical 5'-CACATGT Twist sites)~\cite{pmid20981027,pmid15026577}.
However, our DC detector is different than a detector built strictly from the $\mathcal D_{\textbf{DC}}$ data set (the set of all 12 orthologs for each {\em melanogaster} locus), since we included additional ortholog CRMs of the NEEs.  
 
Furthermore, within the NEEs one could imagine that the spacer has diverged in species that we analyzed that were not analyzed previously, and our choice of the spacer window is an interval not the same as previous choices. For example, Papatsenko et al.~\cite{pmid15795372,pmid19651877} showed that binning the spacers between Dorsal and Twist that there were various optimal bins (namely 14bp, 20bp, and 53bp). It is also possible that the spacer defining the distance of Dorsal and Twist in {\em D. melanogaster} has further diverged in its ortholog species, in particular those not previously analyzed and annotated. 

Szymanski \textit{et al.}\cite{pmid7774581} used DU-like Dorsal sites in his systematic study of the role spacing has between Dorsal and Twist site, suggesting that Dorsal Twist still cooperate if one uses a DU-like binding site, which is further corroborated by systematic studies from Fakhouri \textit{et al.}\cite{pmid20087339} that also used A-track Dorsal sites for the primary Dorsal sites.  These studies suggest evolution could have fixed either a DC or a DU type site at an NEE locus utilizing Dorsal Twist linked sites for synergy, which would deteriorate our claim that DC and DU are really different types of Dorsal sites.  However, it is highly unlikely that all these sites would have fixed with the same sequence unless they were functional or else if the CRMs containing them were duplications.   
\subsection*{The OR gate and the CB detector}

The OR gate scores any input k-mer with both conditional detectors DC and DU, and then outputs simply the lowest energy score.  Similar detectors have been represented in the literature as a Hidden Markov Model or as a mixture model~\cite{pmid16236723,Hannenhalli01062005}.
Each component of the mixture is simply a conditional PWM, where the mixing frequencies are estimated as the fraction of training data that is associated with a particular component (or class) of the mixture.  The mixture is defined as:
  \begin{equation}\label{mixpwm}
P(S)=\sum_{c} \frac{\exp{-E(S|C=c})}{Z_c} P(C=c),
\end{equation}
where $E(S|C)$ is in units of $\lambda_1$ (which is further assumed to have been calibrated to thermal energy units), and $Z_c=\sum_{S\in \mathcal S} \exp{E(S|C=c)}$, where $\mathcal S$ is the set all possible k-mers, $|\mathcal S|=4^k$.

The CB detector is the traditional position independent probability model (PWM) of binding sites, where the PWM is constructed by aligning all of the sites in the $\mathcal D_{\textbf{CB}}$  data simultaneously.
Recall from \Eref{pwm1}
\begin{equation}\label{cbpwm}
P(S)=\prod_i P(S_i).
\end{equation}
where, as a consequence of Bayes' Theorem $P(S_i)=\sum_c P(S_i|C=c)P(C=c)$.
However, for a sequence of bases, $\prod_i P(S_i)=\prod_i \sum_c P(S_i|C=c)P(C=c) \neq \sum_{c}\prod_i P(S_i|C=c)P(C=c)$, where the last expression is the mixture of \Eref{mixpwm}, and is equivalent to a marginalization of the sequence over the classes.  The mixture distribution of the sequence over classes can only be factorized as a product of position distributions {\em given} the class. 
We justify our approximation of the marginal sequence distribution over classes as a PWM (the CB PWM) in the Supplement section 1.8. 

 The mixture model was used by Hannenhali et al.~\cite{Hannenhalli01062005} in a similar form as the OR gate, where a given transcription factor's binding preference was described by two PWMs.  There the authors scanned a given CRM or promoter with both PWMs and selected the highest scoring sites as hits, where the threshold for a hit was determined by the mixing frequencies--the proportion of known sites that are used in constructing each PWM.  Upon scoring all the sites within their promoters, the scores were ranked for a given PWM, and then the fraction of sites equal to the mixing frequency were considered positives.  This method is different than the OR gate presented here in that we do not use the mixing frequencies in discriminating Dorsal binding sites from background DNA.  The OR gate discriminates sites from non-sites by checking if the minimum (i.e., best) score of the component detectors is below the energy threshold.  By always choosing the lowest energy score among the given components as the detector's overall energy score, the benefit of an increased True Positive Rate of the detector is partially cancelled by the cost of an increased False Positive Rate.  However, this cost is only in effect at high energies (non-specific sites), where it is unlikely that evolution or physical binding is having any functional effect on the organism.  Hence, the OR gate is a useful model for increased sensitivity in the low energy regime.

 \subsection*{Information that detectors have about Dorsal binding sites}
In a physical NVE ensemble (fixed particle number $N$, fixed volume $V$, fixed energy $E$) the information content of the distribution of momentum and positions (the distribution function) is conserved.  This means the number of bits necessary to store the position and momentum information is conserved in time relative to the maximum storage capacity defined by a lattice over phase space (the space of coordinates).  For example, if the distribution function is a uniform distribution over phase space, it has zero information content.  

Similarly, evolutionary systems under adaptive maintenance (purifying selection) conserve information stored in their genes~\cite{pmid22320231}.  The inheritance of information implies that parents pass a fixed number of bits to their progeny.  And just as in the NVE ensemble where coordinates and momentum are not conserved, similarly in evolution sequences are variable, but the sequence's  information content is conserved.  However, when the fixed energy constraint of the NVE system is relaxed and the system exchanges energy with a much larger environment, the system's original information content may deteriorate until the system equilibrates with its surroundings. Biological systems harness energy from their environment to maintain their information content in the never-ending fight against the second law~\cite{Adami2002,Carothersetal2004}.

 The mutual information between sequences and the OR gate's predictions in \Fref{rocfig}  suggests that the conditional distributions of functional Dorsal binding sites have encoded synergistic and antagonistic information about flanking sequence features (presence of Twist) that causes the likelihood to correctly predict the presence of Dorsal to shift downwards in energy (as observed by the shift of the mutual information of the OR gate relative to CB in \Fref{rocfig}).  This shift may have been a necessary adaptation in the way Dorsal regulates its targets.  For example it is possible that at the phylum level, possibly before the neuroectoderm evolved, Dorsal only needed to regulate the mesoderm and ectoderm. When the neuroectoderm evolved, Dorsal evolved the ability to recognize two subtypes of binding site ensembles, a function that would help to resolve the neuroectoderm Dorsal targets from the more ancient germ layers (mesoderm and ectoderm).  In this sense, Dorsal's adaptation to its local environment is seen as the shifted mutual information relative to the CB detector (which just treats all binding loci identically).  Dorsal could then use this information to its advantage, in Dorsal real time so to speak, to make better decisions about binding.    
  
 The shift in the mutual information plot in  \Fref{rocfig}B is not as visible in the ROC curves in \Fref{rocfig}A, in which we used the same TPR and FPR for the detectors.  This is because, in general, energy level spacing is not accounted for in an ROC curve, implying that detectors with similarly ranked sequences may actually have different spacings between their energy levels, and the minimum energies of the scales may be shifted relative to one another.  For example, DC's ground state is below CB's ground state, which is why the OR gate contains some information at negative energy (as DC's ground state is at about -0.8 in energy units as seen from the horizontal axis of \Fref{rocfig}).
     
 The degree to which the OR gate's ROC does appear shifted relative to CB's ROC in \Fref{rocfig}A is partly due to the fact that the ranking of sequences of the DC detector and DU detector is very similar; it is the energy level spacing that is dramatically different between the conditional detectors.  For example, using a substitution model that penalizes all mismatches from the consensus sequence with the same energy score (see the Appendix of Ref.~\cite{pmid3612791} for details) leads to the elegant formula that a consensus base occurs with probability $1-\frac{m}{3k}$, and that an error or substitution occurs with probability $\frac{m}{3k}$, where \textit{k} is the length of the sequence and \textit{m} is the number of mismatches from the consensus (the 3 in the denominator is due to the three ways a mismatch from a consensus DNA nucleotide can occur).  Weak sites will be seen to have large \textit{m}, which to a degree can be seen as the DU training data. Similarly, strong sites will have small \textit{m}, which can be seen as the DC data.  Hence in this substitution model, the difference between DC and DU is not in the ordering of their ranked sequences, rather the difference lies in their energy level spacings (which can be seen by changing $m$ which affects the energy spacing formula \Eref{epwm2}).  
 
 This picture of DC functional sequences being a strong version of DU's sequences is consistent with our findings that their median energies differed by almost two units, and with Papatsenko \textit{et al.}'s findings~\cite{pmid15795372} that Dorsal binding sites necessary in limiting concentrations of Dorsal protein (such as in the neuroectoderm) tend to have higher information scores (lower energy scores), than other Dorsal sites such as sites active in the mesoderm~\cite{pmid15795372}.  It is also consistent with the mathematical definition of ``specialized" sites from Erives \textit{et al.}\cite{pmid15026577} and the $D_{\beta}$ sites of Crocker \textit{et al.}~\cite{pmid20981027} who defined these sites based on how they were detected (similar to MEME's One Occurrence Per Sequence setting (OOPS)\cite{pmid7584439}, the specialized sites were one site per NEE CRM sequence, where each discovered site shared the highest sequence similarity between the selected sites between the CRMs), which in a sense, is the Dorsal site that had the slowest mutation rate (i.e., under the strongest purifying selection).       
\subsection*{Conditional detectors}

In \Fref{MiSP} we see that the DC detector can resolve whether a Twist site is in the spacer window or not if the detector fires when $E(S|C) < 3$ (see \Eref{cepwm}). The resolution is not perfect in this regime: the DC detector still has an error rate, which we define as $1-2^{-H(\mathcal C| \mathcal P)}$, where the conditional entropy is defined as:
\begin{equation}\label{conduncer}
H(\mathcal C| \mathcal P)=H(C)-I(\mathcal C; \mathcal P).
\end{equation}
The conditional entropy, $H(\mathcal C| \mathcal P)$, is simply the uncertainty of $\mathcal C$ given $\mathcal P$.  But what does this mean for a DC detector?  We interpreted this conditional uncertainty as a measure of the detector's uncertainty about the underlying Dorsal binding site sequence given how well it predicted its context.  For example, if we assume $H(\mathcal C)=1$ bit while DC's information is $I(\mathcal C; \mathcal P)=0.3$, then plugging into \Eref{conduncer} we have
\begin{equation}
H(\mathcal C| \mathcal P)=1-0.3 = 0.7\ {\rm bits},
\end{equation}
and hence Dorsal has decreased its uncertainty about its context.  

If the mutual information $I(\mathcal C; \mathcal P)$ was maximal (1 bit), then Dorsal could predict with perfect accuracy whether Twist was proximal or distal.  At the opposite extreme where the Dorsal detector does no better than random guessing, we see that it would take about two guesses on average to predict if Twist will be near a binding site sequence.  From an evolutionary point of view, the information $I(\mathcal C; \mathcal P)$ encoded in Dorsal binding sites can be seen as a message passed from an ancestral population of flies to its descendants.  Here, the message instructs Dorsal to interact with Twist, and is encoded in the DNA of Dorsal binding sites.

\section*{Conclusion}

Position Weight Matrices represent a linear coarse-grained physical lattice model of DNA-transcription factor binding.  At the DNA sequence level and at the level of Darwinian selection PWMs represent one of simplest possible linear models.  In the case that each position within a binding site is independently interacting with the protein binding domain, it makes sense to use a simple model for binding since the affinity (the phenotype) is linear, and hence natural selection may behave as if a linear model.  However, binding site sequences may be dependent, and hence linear models will miss important information.  By conditioning PWMs based on the variables that are causing the dependency structure within binding sites it is possible to resolve the binding sites into independent classes that can then each be modeled as conditionally independent PWMs.

The necessity of introducing nonlinear sequence models into binding site sequence models is known to help improve binding site sequence detection, and to give a more realistic perspective to binding site models.  A number of groups have introduced similar models for discovery of co-occuring motifs~\cite{liu2001bioprospector,pmid21486752,pmid16873468,pmid14762058},\cite{barasch,pmid18426806,pmid19286833,GuhaThakurta01072001,gadempmid19193149,Moses04phylogeneticmotif}.  In addition, others have looked at the influence of symmetries in the flanking sequence of binding sites~\cite{pmid21723826,pmid25313048}.  Here we placed our analysis in the context of Berg and von Hippel's population genetics model that is related to thermodynamics, and hence the interaction term could be placed inside of thermodynamics occupancy models of transcription factors.

Our conditional PWMs account for epistatic interactions between Dorsal binding sites and their \textit{cis}-context.  We showed that Dorsal binding sites contain on average around 0.5 bits of information about the presence of Twist in the flanking sequence of each Dorsal site (see Table 1), thereby contributing to disentangling the dependency structure of Dorsal binding sites active in fly development.  In the future, our model can be incorporated in the annotation of binding sites of regulatory regions, and could be used for modeling cooperativity and antagonistic interactions directly from the sequence level. Such models could be used by occupancy models of transcription factors that predict gene expression, such as those in Refs.~\cite{pmid20862354,pmid20087339}. 

\section*{Acknowledgements} We would like to thank David Arnosti and C. Titus Brown for extensive discussions, as well as the members of the Adami Lab. This work was supported in part by NSF's BEACON Center for the Study of Evolution in Action, under Contract No. DBI-0939454.

\section*{References}
\providecommand{\newblock}{}


\providecommand{\newblock}{}
\begin{thebibliography}{10}
\expandafter\ifx\csname url\endcsname\relax
  \def\url#1{{\tt #1}}\fi
\expandafter\ifx\csname urlprefix\endcsname\relax\def\urlprefix{URL }\fi
\providecommand{\eprint}[2][]{\url{#2}}

\bibitem{landaumech}
Landau D and Lifshitz I 1976 {\em {Mechanics}\/} vol~1 (Butterworth Heinemann)

\bibitem{pmid19104053}
Davidson E~H and Levine M~S 2008 {\em Proc. Natl. Acad. Sci. U.S.A.\/} {\bf
  105} 20063--20066

\bibitem{pmid9169833}
Arnone M~I and Davidson E~H 1997 {\em Development\/} {\bf 124} 1851--1864

\bibitem{pmid17903288}
Lassig M 2007 {\em BMC Bioinformatics\/} {\bf 8 Suppl 6} S7

\bibitem{pmid10892643}
Carroll S~B 2000 {\em Cell\/} {\bf 101} 577--580

\bibitem{pmid12777501}
Wray G~A, Hahn M~W, Abouheif E, Balhoff J~P, Pizer M, Rockman M~V, Romano L~A
  and Wray G~A 2003 {\em Mol. Biol. Evol.\/} {\bf 20} 1377--1419

\bibitem{pmid9581503}
Stormo G~D and Fields D~S 1998 {\em Trends Biochem. Sci.\/} {\bf 23} 109--113

\bibitem{pmid8080080}
Fields D~S and Stormo G~D 1994 {\em Anal. Biochem.\/} {\bf 219} 230--239

\bibitem{hill}
Hill T~L 1985 {\em Cooperativity Theory in Biochemistry: Steady-state and
  Equilibrium Systems\/} (New York: Springer-Verlag)

\bibitem{pmid20877328}
Stormo G~D and Zhao Y 2010 {\em Nat. Rev. Genet.\/} {\bf 11} 751--760

\bibitem{pmid9268651}
Fields D~S, He Y, Al-Uzri A~Y and Stormo G~D 1997 {\em J. Mol. Biol.\/} {\bf
  271} 178--194

\bibitem{pmid3612791}
Berg O~G and von Hippel P~H 1987 {\em J. Mol. Biol.\/} {\bf 193} 723--750

\bibitem{pmid15511292}
Sinha S, Blanchette M and Tompa M 2004 {\em BMC Bioinformatics\/} {\bf 5} 170

\bibitem{Berg15081992}
{Berg, O G} {1992} {\em {Proceedings of the National Academy of Sciences}\/}
  {\bf {89}} {7501--7505}

\bibitem{hobson}
Hobson A 1971 {\em {Concepts in Statistical Mechanics}\/} (New York: Gordon and
  Breach)

\bibitem{pmid22887818}
Stewart A~J, Hannenhalli S and Plotkin J~B 2012 {\em Genetics\/} {\bf 192}
  973--85

\bibitem{pmid21723826}
Sela I and Lukatsky D~B 2011 {\em Biophys. J.\/} {\bf 101} 160--166

\bibitem{345660479}
von Hippel P~H 1979 On the molecular bases of the specificity of interaction of
  transcriptional proteins with genome dna {\em Biological Regulation and
  Development\/} vol~1 ed RFGoldberger (New York: Plenum Publishing) pp
  279--347

\bibitem{Atkins}
Atkins P and de~Paula J 2002 {\em Physical Chemistry\/} (W.H. Freeman and
  Company)

\bibitem{pmid7317363}
Berg O~G, Winter R~B and von Hippel P~H 1981 {\em Biochemistry\/} {\bf 20}
  6929--6948

\bibitem{pmid25313048}
Afek A, Schipper J~L, Horton J, Gordan R and Lukatsky D~B 2014 {\em Proc. Natl.
  Acad. Sci. U.S.A.\/} {\bf 111} 17140--17145

\bibitem{pmid14983022}
Brown C~T and Callan C~G 2004 {\em Proc. Natl. Acad. Sci. U.S.A.\/} {\bf 101}
  2404--2409

\bibitem{Schneider.Stephens1990}
Schneider T~D and Stephens R~M 1990 {\em Nucleic Acids Res.\/} {\bf 18}
  6097--6100

\bibitem{Gehring1998}
Gehring W~J 1998 {\em Master Control Genes in Development and Evolution: The
  Homeobox Story\/} (New Haven, CT: Yale University Press)

\bibitem{Davidson2001}
Davidson E~H 2001 {\em Genomic Regulatory Systems: Development and Evolution\/}
  (San Diego, CA: Academic Press)

\bibitem{Davidson2006}
Davidson E~H 2006 {\em The Regulatory Genome: Gene Regulatory Networks in
  Drvelopment and Evolution\/} (San Diego, CA: Academic Press)

\bibitem{Brown2008}
Brown C~T 2008 {\em Methods in cell biology\/} {\bf 87} 337--365

\bibitem{pmid20981027}
Crocker J, Potter N and Erives A 2010 {\em Nat Commun\/} {\bf 1} 99

\bibitem{bialek}
Bialek W 2012 {\em Biophysics Searching for Principles\/} (Princeton University
  Press)

\bibitem{pmid20339533}
Siddharthan R 2010 {\em PLoS ONE\/} {\bf 5} e9722

\bibitem{pmid18725950}
Sharon E, Lubliner S and Segal E 2008 {\em PLoS Comput. Biol.\/} {\bf 4}
  e1000154

\bibitem{pmid15315758}
Leung T~H, Hoffmann A and Baltimore D 2004 {\em Cell\/} {\bf 118} 453--464

\bibitem{pmid19372434}
Meijsing S~H, Pufall M~A, So A~Y, Bates D~L, Chen L and Yamamoto K~R 2009 {\em
  Science\/} {\bf 324} 407--410

\bibitem{busse}
Busse M~S, Arnold C~P, Towb P, Katrivesis J and Wasserman S~A 2007 {\em The
  EMBO Journal\/} {\bf 26} 3826--3835

\bibitem{pmid21890896}
Mrinal N, Tomar A and Nagaraju J 2011 {\em Nucleic Acids Res.\/} {\bf 39}
  9574--9591

\bibitem{lawr}
Lawrence P 1992 {\em {{T}he {M}aking of a {F}ly}\/} 1st ed (Wiley-Blackwell)

\bibitem{Hong23122008}
Hong J~W, Hendrix D~A, Papatsenko D and Levine M~S 2008 {\em Proceedings of the
  National Academy of Sciences\/} {\bf 105} 20072--20076

\bibitem{pmid15788537}
Levine M and Davidson E~H 2005 {\em Proc. Natl. Acad. Sci. U.S.A.\/} {\bf 102}
  4936--4942

\bibitem{pmid19843594}
Perry M~W, Cande J~D, Boettiger A~N and Levine M 2009 {\em Cold Spring Harb.
  Symp. Quant. Biol.\/} {\bf 74} 275--279

\bibitem{pmid16271864}
Moussian B and Roth S 2005 {\em Curr. Biol.\/} {\bf 15} R887--899

\bibitem{pmid16198288}
Stathopoulos A and Levine M 2005 {\em Dev. Cell\/} {\bf 9} 449--462

\bibitem{pmid17322397}
Zeitlinger J, Zinzen R~P, Stark A, Kellis M, Zhang H, Young R~A and Levine M
  2007 {\em Genes Dev.\/} {\bf 21} 385--390

\bibitem{pmid18986212}
Crocker J, Tamori Y and Erives A 2008 {\em PLoS Biol.\/} {\bf 6} e263

\bibitem{pmid15026577}
Erives A and Levine M 2004 {\em Proc. Natl. Acad. Sci. U.S.A.\/} {\bf 101}
  3851--3856

\bibitem{pmid15128669}
Markstein M, Zinzen R, Markstein P, Yee K~P, Erives A, Stathopoulos A and
  Levine M 2004 {\em Development\/} {\bf 131} 2387--2394

\bibitem{pmid1925551}
Kosman D, Ip Y~T, Levine M and Arora K 1991 {\em Science\/} {\bf 254} 118--122

\bibitem{pmid16750631}
Zinzen R~P, Senger K, Levine M and Papatsenko D 2006 {\em Curr. Biol.\/} {\bf
  16} 1358--1365

\bibitem{pmid7774581}
Szymanski P and Levine M 1995 {\em EMBO J.\/} {\bf 14} 2229--2238

\bibitem{pmid8453668}
Jiang J and Levine M 1993 {\em Cell\/} {\bf 72} 741--752

\bibitem{pmid1325394}
Ip Y~T, Park R~E, Kosman D, Bier E and Levine M 1992 {\em Genes Dev.\/} {\bf 6}
  1728--1739

\bibitem{Gallo21102010}
Gallo S~M, Gerrard D~T, Miner D, Simich M, Des~Soye B, Bergman C~M and Halfon
  M~S 2010 {\em Nucleic Acids Research\/}

\bibitem{pmid19651877}
Papatsenko D, Goltsev Y and Levine M 2009 {\em Nucleic Acids Res.\/} {\bf 37}
  5665--5677

\bibitem{pmid2082934}
Berg O~G 1990 {\em Biomed. Biochim. Acta\/} {\bf 49} 963--975

\bibitem{barasch}
Barash Y, Elidan G, Kaplan T and Friedman 2003 Modeling dependencies in
  protein-{DNA} binding sites {\em Proc of the 7th Ann Int Conf in Comp Mol Bio
  (RECOMB)\/} pp 28--37

\bibitem{pmid1582412}
Pan D and Courey A~J 1992 {\em EMBO J.\/} {\bf 11} 1837--1842

\bibitem{pmid15795372}
Papatsenko D and Levine M 2005 {\em Proc. Natl. Acad. Sci. U.S.A.\/} {\bf 102}
  4966--4971

\bibitem{pmid20087339}
Fakhouri W~D, Ay A, Sayal R, Dresch J, Dayringer E and Arnosti D~N 2010 {\em
  Mol. Syst. Biol.\/} {\bf 6} 341

\bibitem{pmid16236723}
Mustonen V and Lassig M 2005 {\em Proc. Natl. Acad. Sci. U.S.A.\/} {\bf 102}
  15936--15941

\bibitem{Hannenhalli01062005}
Hannenhalli S and Wang L~S 2005 {\em Bioinformatics\/} {\bf 21} i204--i212

\bibitem{pmid22320231}
Adami C 2012 {\em Ann. N. Y. Acad. Sci.\/} {\bf 1256} 49--65

\bibitem{Adami2002}
Adami C 2002 {\em {BioEssays}\/} {\bf 24} 1085--1094 ISSN 1521-1878

\bibitem{Carothersetal2004}
Carothers J~M, Oestreich S~C, Davis J~H and Szostak J~W 2004 {\em J. American
  Chem. Society\/} {\bf 126} 5130--5137

\bibitem{pmid7584439}
Bailey T~L and Elkan C 1995 {\em Proc Int Conf Intell Syst Mol Biol\/} {\bf 3}
  21--29

\bibitem{liu2001bioprospector}
Liu X, Brutlag D, Liu J {\em et~al.\/} 2001 Bioprospector: {D}iscovering
  conserved {DNA} motifs in upstream regulatory regions of co-expressed genes
  {\em Pac Symp Biocomput\/} vol~6 pp 127--138

\bibitem{pmid21486752}
Bais A~S, Kaminski N and Benos P~V 2011 {\em Nucleic Acids Res.\/} {\bf 39} e76

\bibitem{pmid16873468}
Georgi B and Schliep A 2006 {\em Bioinformatics\/} {\bf 22} e166--173

\bibitem{pmid14762058}
Bulyk M~L, McGuire A~M, Masuda N and Church G~M 2004 {\em Genome Res.\/} {\bf
  14} 201--208

\bibitem{pmid18426806}
Hannenhalli S 2008 {\em Bioinformatics\/} {\bf 24} 1325--1331

\bibitem{pmid19286833}
Pape U~J, Klein H and Vingron M 2009 {\em Bioinformatics\/} {\bf 25} 2103--2109

\bibitem{GuhaThakurta01072001}
GuhaThakurta D and Stormo G~D 2001 {\em Bioinformatics\/} {\bf 17} 608--621

\bibitem{gadempmid19193149}
Li L 2009 {\em J. Comput. Biol.\/} {\bf 16} 317--329

\bibitem{Moses04phylogeneticmotif}
Moses A~M and Eisen M~B 2004 {\em Pac. Symp. Biocomput\/} {\bf 324} 324--335

\bibitem{pmid20862354}
He X, Samee M~A, Blatti C and Sinha S 2010 {\em PLoS Comput. Biol.\/} {\bf 6}

\bibitem{pmid9021275}
Galtier N, Gouy M and Gautier C 1996 {\em Comput. Appl. Biosci.\/} {\bf 12}
  543--548

\bibitem{BSA}
Durbin R, Eddy S, Krogh A and Mitchison G 1998 {\em Cambridge Press\/}

\bibitem{pmid7584402}
Bailey T~L and Elkan C 1994 {\em Proc Int Conf Intell Syst Mol Biol\/} {\bf 2}
  28--36

\bibitem{lawrence1993detecting}
Lawrence C, Altschul S, Boguski M, Liu J, Neuwald A, Wootton J {\em et~al.\/}
  1993 {\em Science\/} {\bf 262} 208--208

\bibitem{pmid22683811}
Obbard D~J, Maclennan J, Kim K~W, Rambaut A, O'Grady P~M and Jiggins F~M 2012
  {\em Mol. Biol. Evol.\/} {\bf 29} 3459--3473

\bibitem{10.1371}
Habib N, Kaplan T, Margalit H and Friedman N 2008 {\em PLoS Comput Biol\/} {\bf
  4} e1000010

\bibitem{Down01012005}
Down T~A and Hubbard T~J~P 2005 {\em Nucleic Acids Research\/} {\bf 33}
  1445--1453

\bibitem{Jain:1999:DCR:331499.331504}
Jain A~K, Murty M~N and Flynn P~J 1999 {\em ACM Comput. Surv.\/} {\bf 31}
  264--323

\bibitem{pmid17478497}
Mahony S and Benos P~V 2007 {\em Nucleic Acids Res.\/} {\bf 35} W253--258

\bibitem{Nemenman02entropyand}
Nemenman I, Shafee F and Bialek W 2002 Entropy and inference, revisited {\em
  Advances in Neural Information Processing Systems 14\/} (MIT Press) pp
  471--478

\bibitem{MEP}
Levine R~D and Tribus M (eds) 1978 {\em {The Maximum Entropy Formalism}\/}
  (Cambridge, MA: MIT Press)

\bibitem{bishop}
Bishop 1985 {\em Pattern Recognition and {Machine Learning}\/} vol~1
  (Rockville, MD: Computer Science Press)

\end{thebibliography}

\setcounter{section}{1}
\setcounter{table}{1}
\section*{Methods Supplement}

\subsection{ Alignment of cis-regulatory modules and collection of $\mathcal D_{\textbf{CB}}$} \label{muscle}
	We used the sequence editor SEAVIEW's default MUSCLE Multiple Sequence Alignment settings \cite{pmid9021275} to align a given gene's 12 orthologous CRMs~\cite{BSA}.  Given the alignment we then manually extracted the blocks that contained the known $\textit{D. mel}$ binding sites, which allowed for flanking sequence to be extracted (these blocks on average spanned about 15bp with no gaps across the 12 species).  The data set of all 12 extracted orthologs (sometimes less than 12 if a binding site was not in the block) for all the $\textit{D. mel}$ Dorsal binding site loci is labelled as $\mathcal D_{\textbf{CB}}$.  The $\mathcal D_{\textbf{CB}}$ data set is in our available upon request, along with a Fasta file of the $\mathcal D_{\textbf{DU}\textit{mel}}$ and $\mathcal D_{\textbf{DC}\textit{mel}}$ sites that contains Pubmed ID of at least one paper that verified the binding site, the REDFLY record number of each site (if it was footprinted), and the primary author of the Pubmed article.

\subsection{GEMSTAT modifications for locus annotation of CRMs}\label{knownloc}
	 Given the known sites, $\mathcal D_{\textbf{CB}}$, with their flanking sequence from the block alignments we then created an exact search algorithm that allowed us to estimate the coordinate of the known site within the regulatory region for data structures used by GEMSTAT, a platform developed in the Sinha lab~\cite{pmid20862354}.  This program has a various inputs, which are irrelevant for our current purposes, the relevant input is a set of PWMs that represent transcription factors, and the set of CRMs regulated by these factor's motifs (PWMs).  We extended the GEMSTAT input to allow for a raw Fasta set of variable length binding sites (namely our Fasta file for the data set $\mathcal D_{\textbf{CB}}$).  Each of these sites and its reverse complement was transformed to a probability PWM ``singleton" representation (the probability is equal to one for the observed nucleotide and zero for the other nucleotides).  The longest binding site of length n in the Fasta file determined the length of all the singletons.  Binding sites in the Fasta file that were of length k, where $k < n$, then had $d=n-k$ columns of zeros 'padding' the last d columns of their singleton PWM.
	 
	 We then constructed a distance matrix from the singletons, where each singleton is represented by a row and column in the matrix.   The matrix elements of the distance matrix were defined as a normalized euclidean dot product between the singletons (where corresponding components of each singleton multiple each other).  The normalized dot product between the singleton $S^x$ in row x and singleton $S^y$ in column y is defined as sum over all the element-wise multiplications:
\begin{equation}
	S^x \bullet S^y   = \frac{\sum_{ij}^{nn} S_{ij}^xS_{ij}^y}{\sum_{ij}^{nn}S_{ij}^xS_{ij}^x}.
	 \end{equation}	 
	 The row singletons are used for normalization, and hence for a given row x, by iterating over all columns we can filter out any identical singletons.
	 
For example, if we have 400 positive strand binding sites in our Fasta file, we will have a 800x800 distance matrix (400 of the rows corresponding to positive stranded sites, and another 400 rows for the negative strands); where the matrix elements are normalized Euclidean dot products between any two singletons.  By screening the distance matrix for elements that were equal to one (which stands for a duplicate sequence or possible symmetric sites) we are able to determine which singletons are unique.  Asymmetric sites contain an instance for each strand (which means they are not unique), which is accounted for in our ``overlapping" sites processing step.  
	 
Collecting all the unique singletons, we then annotate (scan) the CRMs with each unique singleton using exact match (\textit{e.g.}, zero energy singleton PWM threshold).  This step allows us to map each binding site in our Fasta file to a unique locus within the CRMs, thereby attaining the coordinates of all our known binding sites within the CRM coordinate system, which is an essential step in the spacer calculation of \Eref{cluster}.  
	 
Each unique singleton has a personal factor identification (a name), which we mutate to the name 'Dorsal' for every singleton.  Associating the name `Dorsal' to the annotated sites within the CRMs is a necessary step in order to compute the spacer between each of the annotated `Dorsal' sites and any predicted motifs sites within each CRM.  The coordinate defines the start or end of a binding site depending on what strand of the CRM was annotated as a positive hit.  For asymmetric known sites that match the bottom strand, the coordinate defines the end of the binding site, while matches on the positive (or `top') strand of the CRM indicate the start of the site.  The asymmetric sites are further processed to cull out any sites that overlap at a specific locus, where the process is described below.    

	   \subsection{Overlapping site processing}\label{overlap}
  In order to create independent loci, we wanted to have only one hit per binding site, so we culled all overlapping sites and overlapping footprints.  In the singleton construction of known binding sites our set of unique singletons frequently contains asymmetric binding sequences which means both the top and bottom strand sequence at a particular locus will have an associated singleton (overlapping binding sites).  We are able to choose just one representative for a given locus by an algorithm explained in the `Best Predictions' section below.     
 
	   \subsection{Error In estimating the spacer length between known Dorsal loci and Twist sites}
	The spacer between two annotated binding sites is determined by the coordinates of the start site of each binding site using the CRM coordinate system.  However, 'known' Dorsal sites were annotated in the CRMs using Euclidean dot product search algorithm (discussed above in the section `GEMSTAT Modifications For Locus Annotation Of CRMs'), which has an uncertainty in start site of a binding site due to the searching singleton PWM being longer than the actual binding site (due to the flanking sequence).  
	We have about a 6 base-pair error in the exact spacer length for our length 9 binding sites of Dorsal (assuming the Twist site or other potential cooperating site annotation uses a correct length PWM).  This is due to the actual Dorsal binding site not being 15 bps (which is the typical length of a site used in our exact search algorithm).  However, in practice, many of the loci had smaller appended flanking sequence, which would reduce the error in the spacer length calculation.  Furthermore, for binding sites that were centered (which they usually are) in the extracted block from MSA, the spacer length error would be reduced in half.  Here we call this spacer length error and not bias, because we simply do not know exactly where the site resides within the blocks extracted from the MSA.  
	Given the coordinates of the known Dorsal binding sites within the \textit{cis}-regulatory module we then defined a PWM for putative cooperating factor with Dorsal (such as a Twist PWM) and set a threshold on this factor's energy to annotate it's predicted sites (where we assume this factor's annotation uses a correct length PWM).
	
	\subsection{PWM Best \textit{predictions} of binding site loci }  
By scanning or scoring each possible subsequence of length $k$ within a CRM with the PWM one can filter out all the subsequences that do not match the PWM, where a match is defined as having an energy score below a defined threshold.  The coordinates of the subsequences that match the PWM relative to the CRM coordinate system can then be used to determine the locus of the predicted binding site.
	
Scanning CRMs with a PWM frequently results in multiple overlapping binding sites due to symmetry (positive and negative strand being called a hit) and due to re-occurring patterns in motifs (such as repeated bases like 5'-AAAA).  In order to have non-overlapping binding sites we processed the set of match sites from the CRM scan to construct a list of non-overlapping sites.  
	  
We treated each position within a CRM as the start site of a binding site of length $k$ that was scored by the PWM using Eq.~(\ref{epwm}) (or Eq.~(\ref{cepwm}) depending on the model being used for predictions).  The reverse complement of each potential binding site was also scored by the PWM.  Each length $k$ sequence (potential binding site) was stored in a data structure, a k-mer, which contained attributes of the potential binding site like the coordinate and strand (relative to the CRM) and the energy score.  The k-mers below the energy threshold were selected as a hit, and temporarily stored in a hit list.  In order to have no overlapping hits we sorted the k-mer list according to energy scores.  The coordinate attribute of the k-mer with the minimum energy, the best site, was used to mask out any overlapping hits.  This best k-mer site was passed to a storage vector, which would ultimately contain the annotated k-mer binding sites of the CRM.  Upon deleting the minimum energy k-mer site along with the masked out k-mers from the hit list, we iterated the above procedure until the hit list was empty, thereby creating a storage vector of non-overlapping predicted k-mer sites that corresponded to maximum scoring binding sites within the CRM.
	
\subsection{Expectation Maximization Alignment}\label{emalg}
	In this paper EM alignment means Expectation Maximization alignment of binding sites.  We use a one site per sequence setting that resembles the MEME~\cite{pmid7584439} EM one site per sequence algorithm~\cite{pmid7584402}.  A Fasta list of sites is passed to the tool, and for each sequence in the list one internal position of the sequence is defined as the starting position of the inferred binding site.  Only one binding site is allowed per sequence from the list passed to the tool, however, for any given sequence both strands are scored by the current value of the PWM, where the highest scoring site's position, regardless of strand, is saved in order to make the alignment.  The output of the alignment is a PWM.  The tool requires setting the length of the desired PWM and the number of iterations of the Expectation Maximization algorithm and the number of iterations of a sampler.  Recall, the MEME EM simplest form of the algorithm, scores each internal position with a current definition of a PWM.  Then upon scoring all sequences and all internal positions of each sequence within the Fasta list, the Maximum score for each sequence, and hence a corresponding position, is determined for the new starting positions of sequences to be are extracted and used to construct a new PWM, this new PWM is the Expected PWM, in the sense that the Maximum Likelihood values of the expected counts are just the counts themselves.  This new PWM is then reiterated upon all the sequences and all their internal positions, thereby iterating through the EM algorithm.  In addition at each step of the EM iteration, the stored position of the start site of each site in each sequence is shifted by one base pair and then the PWM is recalculated to check for phase shifts.  The shifts are check for both forward and backward shifts up to shifts of half the length of the site~\cite{lawrence1993detecting}.  The EM is wrapped inside of a sampler, which allows for a naive global optimization by random starting positions within each binding site sequence being used as the initial conditions that are past to the EM program.  A global variable stores the best PWM, upon each iteration of the sampler, if the EM output PWM has smaller Kullback-Leibler divergence (i.e., information content) then that PWM is thrown out, otherwise global variable of the best PWM is redefined by the current iteration's PWM, and the sampler continues until the specified number of iterations are exhausted.  The Kullback-Leibler divergence of the distribution (the probability PWM) was estimated from the uniform distribution or from a distribution set by a GC content value.
  In addition we implemented an option to weight each homologous sequence in the alignment based on the divergence time estimated from Obbard \textit{et al.}~\cite{pmid22683811}.  However, we have not fully explored the effects of this weighting scheme, and present no results  with this option. 
  
\subsection{The CB data set, merging and dividing clusters of binding sites}
   The mixture distribution of equation of \Eref{mixpwm} implies the data set $\mathcal D_{\textbf{CB}}$  over all loci can not be aligned simultaneously to form an estimate of $P(S)$, as that only makes sense if one is constructing the CB PWM, which assumes no mixture.  However, a priori, one may not know whether their set of binding site loci is a mixture of different types.  To determine if there is a mixture in the data one must decide on how one will align the mixture model, and whether that alignment should be related to the case that one combines all the data indiscriminately to form an alignment for the CB PWM.  Will all the training data over all classes be lumped together to estimate $P(S)$ for CB, and then the conditional probabilities estimated by partitioning their respective set of aligned sites?  This technique is commonly used in the case that one is given a set of \textit{aligned} data, and one wishes to find mixtures within the aligned data.  Alternatively, will the training data be partitioned into the classes, and then each class aligned individually, and then these class-specific alignments simply be 'merged' to form an estimate of $P(S)$ for CB?

This additional complexity is analogous to the decision made in clustering motifs as to whether one wants a top-down approach (start from the root and \textit{partition}), or a bottom-up approach (start from the leaves of the tree and \textit{merge} (i.e. combine)), see for example \cite{10.1371}\cite{Down01012005}\cite{Jain:1999:DCR:331499.331504}.  We presented results for a bottom-up approach that aligns the training data $\mathcal D_{DC}$ and $\mathcal D_{DU}$ separately to estimate each conditional PWM DC and DU, then CB was based on merging the count matrices of the DC and DU data sets.

The top-down approach aligns all the sites together, then partitions out the classes, and from those partitions builds (without further alignment) the conditional probability PWMs, $P(S|C)$.  The bottom-up method is guaranteed to achieve higher Mutual Information than the top-down approach.  This is because the DC alignment will not necessarily be 'in register' with the independent DU alignment, for example the DU alignment may tend to have the first character with more than 0.5 bits of information shifted relative to the DC alignment (i.e. the binding start site of these two set are shifted). This in turn causes their marginalization to have an increased entropy due to mixing alignments out of register, which in turn causes the mutual information to be boosted, since the conditionals are both now substantially different than the marginal due to registration of the start sites.  Based on results not presented, we found that the top-down approach still preserves the overall trend in Mutual Information versus spacer window, although the signal in the [0,30]bp window for distances of known Dorsal sites from the 5'-CAYAGT motif is reduced by about 40\%.

From a model comparison viewpoint, one may assume the strategy should be to align DC, DU and CB all separately, neither taking a top-down or bottom-up approach.  This strategy does not bias either the mixture model (OR gate) or the CB PWM model.  However, from results not presented, we found this has little effect on the logos for a length 9bp alignment.  Albeit, some alignments (in an ensemble of alignments derived from Gibbs sampling) do choose a `T' rich motif, as opposed to an `A' rich motif.  For our logos, in the case that a `T' rich motif was found, we presented the logo derived from the reverse complement of each \textit{aligned} sequence within the training data set so that all logos would be easily visually comparable by inspecting the logos.  For longer than 9bp length alignments of Dorsal binding sites, we found that the Conditional Dorsal motifs frequently were not in 'register', where registration of the start sites of motifs is based on a motif-motif alignment program like STAMP\cite{pmid17478497}, but can also be seen by inspection of the logos (sometimes).  For example, by inspection, the DC motif for a length 11bp alignment may have had the first position in the alignment with greater than 0.25 bits of information content at position one (in a zero based coordinate system), while DU would have its first position in the logo with more than 0.25 bits at position zero (the start of the logo).

From a physical standpoint, it may be that the conditional binding sites do tend to have a shift in their binding start site positions.  For example, Dorsal may bind to :  5'-AAGGAAATTCC in a DC favored environment, while in the DU favored environment it binds: 5'-GGAAATTCCAA.  If the flanking A's really are a signal, then one must conclude the best representation for CB would be: 5'-AAAGGAAATTCCAAA, which is a motif that is 3 nucleotides longer than either conditional.  The bottom up approach to clustering would miss this signal, since it would merge the two classes such that the CB PWM would contain a fraction of `A's at the first position based on the proportion of the DC data relative to the CB data, and another fraction of `G's at the first position due to the DU motif; thereby not only missing some of the signal, but also interfering with the captured portions of the signal.  This particular case shows that trying to compare motifs based on having the starting position of the motif being in register, will potentially truncate a signal for the CB model.  The logos in \Fref{logofig} are in register (the start of the binding sites are the same), which is based on our findings that length 9 alignments are the most reproducible in terms of registration.  Once we had aligned the length 9 binding sites, we then appended the flanking sequence to each already aligned locus, thereby having greater assurance that the CB PWM would not have this type of interference effects.

  However, if one starts with an alignment that has flanking sequence to begin with, (such as an alignment of length 15bp;) then one could try and discover if the \textit{aligned} sites do contain a mixture of motifs without having to worry about the problems associated with choosing a strand (such as the `A' rich strand), or whether the start sites are in register.  Such an approach was used in Figure 2 of Barash et al.~ \cite{barasch}.  However, we found that setting the Gibbs sampler alignment's length parameter highly influences the alignment.  For example, for a length 9bp alignment, the starting position of the alignment may contain maximum information (2 bits), however if one creates a length 15 alignment this signal (the information) is at least conserved, but may spread into the flanking sequence.  This spreading changes the DNA makeup of the logo of various positions.  This is partly due to there being so many more ways to spread out information among the positions of the alignment when one is using an objective function that runs over 15 positions as opposed to 9bp.  For example, a subsequence of 5'-AAA that is completely conserved in a length 3bp  alignment, when allowed to be length 5bp alignment may converge on 5'-AAAAA with the same information content as the 3bp alignment (or slightly larger information content than the length 3bp alignment, while having a smaller per position information content).  This is due to a mixture over loci of the form: 5'-NAAAN, 5'-AAANN, 5'-NNAAA.
  
 \subsection{CB was designed to be an approximation to a mixture}\label{just}
By choosing `A-rich' strands for representations of DC and DU, we were able to create mixture of PWMs that was not an artifact of the strands when it came to calculating the marginal in the mutual information, and when it came to constructing the CB PWM.  To determine just how similar the CB PWM is to the mixture distribution, we can use the fact that the entropy of a mixture distribution must be greater or equal to the fractional entropies of its component distributions, namely:
\begin{equation}
H(P(S)^{\rm mix}) \geq f_{DC} H(P(S)^{DC}) + f_{DU} H(P(S)^{DU}),
\end{equation}
where $H( P(S)^{\rm mix})$ is the entropy of the mixture model of Eq.~\ref{mixpwm}, and $f_{DC}$ is the fraction of loci in the population that were assigned to class DC and $P(S)^{DC}$ is the probability of $S$ calculated from the DC probability PWM, and $f_{DU}$ is the fraction of loci in the population assigned to class DU and $P(S)^{DU}$ is the probability of sequence $S$ calculated from the DU PWM.  Now if $H(P(S)^{CB})$ is similar in magnitude to $H(P(S)^{\rm mix})$ then it would be reasonable to suggest that $E(S|C) =E(S) + w(S,C)$, where $E(S)$ is estimated from the CB energy PWM.\footnote{For a physical mixture, the correct form of the marginalized energy is: 
\begin{equation}
E(S)=\log{(\frac{P_{ref}}{\sum_{c}\prod_i P(S_i|C=c)P(C=c)})},
\end{equation}
where $P_{ref}$ is the probability of the most probable sequence from the mixture model (i.e. in the mixture model joint distribution $P(S)$ takes the form: $\sum_{c}\prod_i P(S_i|C=c)P(C=c)= P(S)$).}.  We found the entropy of the mixture for the spacer window of [0,30]bp was 8.2 bits while the entropy of CB was 8.4 bits (note the entropy of a probability PWM is $2*k-IC$, where $k$ is the length of the motif and IC is its Information Content calculated using a uniform background distribution over sequences, given by the first term in the Lagrangian Eq.(5) in the main text.  Given that the entropies of these distributions are within a couple decibits and inspection of the logos from figure \ref{logofig} suggests the ranking of sequences by the PWMs is preserved between DC, DU, and CB (and hence by the mixture distribution)- the preservation of sequence, of course, breaks down for the 5'-AAATT and 5'-AAAAT cores of DC and DU, notwithstanding, it seems reasonable, for the Dorsal OR gate, to simply use CB as a proxy for the marginal mixture model in the calculation of $E(S)$.  Without this approximation one would have to use a more complicated data structure in order to calculate $E(S)$, such as a look-up table that stores the probability of all $4^k$ sequences. 
 
\subsection{Conditional Distributions}\label{cdhyper}
The above count table provides the basic elements to estimate the marginal probability of the bases over the two classes.  The maximum likelihood (ML) estimate of the counts from the table are the counts themselves.  Furthermore, any function of a ML estimate is itself the ML estimate.  The ML estimate of the marginal counts of B over C is $n_{B}=n_{B1}+n_{B0}$.  The ML estimate of the marginal probability of $B$ is $\frac{n_B}{n}$, where n is sum over all elements of the table ($n=\sum_B n_B$).  ML estimates of the counts and the probabilities enjoy the property that the estimates are unbiased.  However, the ML estimates of the functionals energy and entropy are biased, where the ML estimate of entropy always underestimates the value of the entropy\cite{Nemenman02entropyand}. The bias in the ML estimate of the energy is more complex to analyze, since the energy (as defined in Eq.~\ref{epwm2}) is equal to the entropy plus an additional extreme value variable (where this extreme value variable is from a Gumbell like distribution).  The bias in these estimators could affect our bioinformatic searches based on a cutoff of the energy, and could affect our calculations of information content and mutual information.  Hence we chose a Bayesian approach that uses a hyperparameter $\beta$ to correct for the small sample bias in entropy and energy.  This approach leads to an estimate of the discrete marginal probability of B over C with a Dirichlet prior with a symmetric hyperparameter $\beta$, defined as $P(B)=\frac{n_{B}+\beta}{n+4*\beta}$.  We used the same $\beta$ for all positions of a PWM\cite{MEP}.  Similarly, the conditional distribution of $B$ given $C$ is defined  as $P(B|C)=\frac{n_{BC}+\beta}{\sum_{B{n_{BC}}+4*\beta}}$, where we use the same $\beta$ for all positions of the conditional probability PWM for our estimates of the conditional distribution of $B$ given $C$. 

To estimate the uncertainty in our count estimates, a frequentist may assume a Poisson counting-like process, which has a well-known property that the expected counts for a set number of trials is equal to the variance of the distribution of counts, which is supported up to the set number of trials.  One can then estimate the confidence interval of their estimates of the expected counts and hence the standard error on $P(B)$.  However, from a Bayesian perspective, the normalized counts are simply samples from a probability simplex (the distribution of distributions)\cite{bishop}.  Here one doesn't estimate standard errors on $P(B)$, rather the variance of the distribution over the probability simplex is a measure of the expected spread of $P(B)$ (\textit{i.e.} how much do we expect $P(B)$ to vary from one alignment (sample) to another, in other words, how reliable is our estimate of $P(B)$).  Thinking of each $B$ as a category, then we can use the Dirichlet as our prior distribution over the categorical distribution $P(B)$ (choosing the Dirichlet as the prior preserves the form of the categorical distribution when new information becomes available that we use to update our estimate of $P(B)$).  The Dirichlet prior has an elegant formula for its variance, which we reproduce here for convenience:
\begin{equation}
\sigma_{P(B)}^2=\langle P(B)^2\rangle-\langle P(B)\rangle^2= \frac{\alpha_B (\alpha_0-\alpha_B)}{\alpha_0^2 (\alpha_0+1)},
\end{equation}
where $\alpha_B$ are the concentrations (hyperparameters) for $B=A,C,G,T$, and $\alpha_0=\sum_B \alpha_B$.
After we observe the sample (the alignment), the variance changes because we've gained new information.  We can (as a consequence of 'conjugacy') simply recycle the formula above with a change of variables $\alpha_B' = \alpha_B + n_B \dots \ \alpha_o'=\alpha_o + n $, this leads to the posterior variance:
\begin{equation}
\sigma_{P(B)_{post}}^2=\frac{(\alpha_B +n_B)(\alpha_0+n -\alpha_B-n_B)}{(\alpha_o+n)^2(\alpha_o+n+n_B + 1)}.
\end{equation}
 We chose to use a symmetric hyperparameter, $\beta$, where $\alpha_B = \beta  \ \forall \ B$, which can be thought of as a ``pseudocount".  Berg and von Hippel used the same analysis with the standard maximum entropy prior, which they detail in their appendix~\cite{pmid3612791}.  
\subsection{ Detector energy thresholds, $E_c$}
From a bioinformatic perspective, a detector's conditional PWMs or the CB PWM must test each potential 9-mer in a CRM by making a prediction as to whether the 9-mer is a binding site or random background DNA.  Hence the prediction is a binary classification that labels each 9-mer as positive or negative.  The positive sites indicate the 9-mer's energy is below an energy threshold, $E_c$ (critical energy), while negative sites have 9-mer sequences with energy above the energy threshold.

We define the bioinformatic specificity, $\nu$, as the cardinality of the number n of sequences of length 9 bp that are considered a positive binding site due to their energy being below the critical energy, divided by the cardinality of the total number $N$ of possible sequences of length 9 bp, where $N=4^9$.  Hence the bioinformatic specificity is $\nu =\frac{n}{N}= \sum_{S\in \mathcal S} P(S) \theta(E(S)-E_c) $, where $\mathcal S$ is the set of all possible sequences (i.e,. the set of cardinality $N$), and $\theta$ is the step function, which acts as an indicator variable that has a value of 'one' when $E(S)$ is below the threshold energy of $E_c$ and $\theta$ has a value of zero otherwise.  Once a bioinformatic specificity is set, we use the estimated cumulative distribution function (cdf) of the CB energy over the $4^9$ sequences to calculate the energy threshold that matches a particular value $\nu$ of the cdf, (where we assume these $4^9$ sequences occur based on the background probability).

We naively build the cdf by nested iterations, which allows us to iterate over all possible 9-mers, $N$.  At each iteration we determine the unique 9-mer sequence $S$'s CB energy $E(S)$ (position independent model), and increment the bin of the energy histogram that corresponds to $E(S)$, where the bin widths were 0.1 in arbitrary units.  To map an energy $E$ to a bin, we map each energy to a bin number (bin identification), where the bin number is $\left \lceil{10*E(S)}\right \rceil$ for a 0.1 precision bin width, or simply $\left \lceil{E(S)}\right \rceil$ for a bin width of 1.  For example, for $\nu=10^{-6}$ we expect $n=\nu N$ possible sequences to be below the energy cutoff, we then can rank each sequence in the set of N unique 9-mers based on their energy score, where the nth sequence's energy is $E_c$.

For a given energy PWM each 9-mer $S$ in a CRM is scored as $E(S)+w(S,C)$, where the shift is determined by the PWM that was trained from class specific data ($w=0$ for CB).  For example, if the spacer window is set at $[0,30]bp$ then the DC detector always expects that there is a cooperating site proximal to it, and hence adds $w(S,proximal)$ to the energy $E(S)$ for a given sequence $S$.  All 9-mers that satisfy the constraint $E(S) + w(S,proximal) < E_c$ are considered a positive hit (where overlapping 9-mers that satisfy this constraint are screened so that the best scoring 9-mer is considered the positive site).
\section*{Results Supplement}\label{rsup}

\subsection{Description of rank sum sampling distribution construction}\label{ranks}

	It's possible that any random data set of binding sites that are used to build detectors using our technique would produce p-value's similar to those found between DC and DU detectors.  Hence, our original data set of Dorsal sites $\mathcal D_{\textbf{CB}}$ was randomly partitioned such that half of the sites from it are placed in $D_1$ and the other half of sequences into $D_2$.  Then we build a detector (a model) with PWMs trained from the $D_1$ sequences, and similarly we build another detector from the $D_2$ data.  Then using our formula for conditional energy we compute the conditional energy for each of the $D_1$ sequences.  For example, let $D_1={S_1,S_2,..S_n}$, where the cardinality of $\mathcal D_{\textbf{CB}}$ is 2n.  Then we compute the energy of these sequences:  $D_{1E}=[E(S_1|1) , E( S_2|1) ,..E(S_n|1)]$.  Similarly the data set $D_{2E}$ will be based on the conditional energies for the corresponding sequences in $D_2$.  Once we have the lists of conditional energies, we compute the median energies between $D_{1E}$ and $D_{2E}$, and use the rank sum test to obtain a p-value.

We repeat this procedure 1000 times and bin the p-value.  Hence we create a distribution of p-values for the rank sum test, which can be used to test if our detector's DC and DU have a significant rank sum p-value against the background of p-values from rank sums of random partitions of the data.

\subsection{Logodds ratio test of DC and DU positive hits} 
 For a given detector, defined by the class value \textit{C}=c, each 9-mer, \textit{S}, in a crm is scored as $E(S)+w(S,C=c)$.  For example, the DC detector always expects that there is a cooperating site nearby, and hence adds \textit{w(S,c=1)} to the energy \textit{E(S)} for any given sequence \textit{S}.  All 9-mers that satisfy the constraint $E(S) + w(S,c=1) < E_c$ are considered a positive hit (where overlapping 9-mers that satisfy this constraint are screened so that the best scoring 9-mer is considered the positive site).

All positive hits for a detector are then classified using the same spacer window scheme that was used in constructing the detector itself (\textit{i.e.}[0,30]bp).  All positive hits that contain a Twist site in the spacing window are classified with class tag `proximal', and the Dorsal sites without a Twist site in the spacing window are classified as `distal'.

We constructed a 2x2 contingency table with table elements $n_{M,C}$ that represent the number of predicted Dorsal loci from detector \textit{M} that match the properties of class \textit{C} (e.g., \textit{C}=p indicates the predicted Dorsal locus had a cooccurring Twist site in the spacer window `proximal'). The detector \textit{M} can be considered a random variable with outcome $M$=DC and $M$=DU, and the class, \textit{C}, another random variable.  The table elements for the DC detector are $n_{DC,p},n_{DC,d}$, for the number of predicted sites \textit{n} from the DC detector that were proximal to Twist's motif, and similarly the number distal from Twist.  

\begin{table}[h]
\centering
  \begin{tabular}{ r|c|c| }
\multicolumn{1}{r}{}
 &  \multicolumn{1}{c}{proximal}
 & \multicolumn{1}{c}{distal} \\
\cline{2-3}
$DC$ & $n_{DC,p}$ &  $n_{DC,d}$ \\
\cline{2-3}
$DU$ &  $n_{DU,p}$ & $n_{DU,d}$ \\
\cline{2-3}
\end{tabular}
\caption{Contingency table of DC and DU detector versus the class type distal and proximal.  Elements of the table are the counts from predictions of each detector for a given energy cutoff and spacer cutoff in a given set of CRMs.}\label{tabless3s}
\end{table}

 The logodds ratio is the ratio of two odds; odds of DC proximal, labelled as ODCp= $\frac{P(n_p|DC)}{P(n_d|DC}$ and odds of DU proximal labelled as ODUp = $\frac{P(n_p|DU)}{P(n_d|DU)}$, where the conditional probabilities (such as $P(n_p|DC)$) are estimated from the 2x2 contingency table in \Tref{tabless3s} with a pseudocount of value one.  The logodds ratio is: $\ln(\frac{ODCp}{ODUp})$.  The sampling distribution of the logodds ratio is a normal distribution with mean zero and width equal to the standard error of the mean.  We are interested in the one-sided test, hence, our p-value estimate is the integral of the sampling distribution from the given logodds ratio to positive infinity (the chances of seeing the value found from the test or a larger value).

 \subsection{Mutual Information between known class tags and the conditional detector's predictions of class tags}\label{misup}
 
We use the mutual information between the binary class variable $\mathcal C$ (`proximal' and `distal', defined by the spacer window) from our known binding sites and the detector's binary prediction $\mathcal P$ of the class to determine the detector's performance at resolving class types.  Here a prediction still means that the detector is testing whether a k-mer is a Dorsal binding site, however, we are additionally checking to see if the binding site locus of the k-mer being tested has the correct flanking sequence feature.

  We use the information identity to transform mutual information into an entropic form:
\begin{equation}  
   I(\mathcal C;\mathcal P)=H(\mathcal P)-H(\mathcal P|\mathcal C)\;.
\end{equation}   
     However, this quantity can only be calculated given a model $M$ (a conditional detector and its corresponding energy threshold), hence, in our own notation we will write this mutual information as $I(\mathcal C;\mathcal P,M=m)=H(\mathcal P,M=m)-H(\mathcal P|\mathcal C,M=m)$, where we make explicit that we know the detector $M$.
     
The variables $\mathcal C, \mathcal P, M$ are all Bernoulli-like.   The value $\mathcal C=0$ indicates class type `distal' for a given binding site, and $\mathcal C=1$ indicates class type `proximal' for a given binding site.  The value $\mathcal P=0$ indicates the detector {\em predicted} the class of a given binding site as `distal', and the value $\mathcal P=1$ indicates the detector {\em predicted} the class of a given binding site as 'proximal'.  The variable $M$'s domain is $M$=DC and $M$=DU.  
     
The entropy $H(\mathcal P,M=m)$ is the entropy of the predicted class distribution, where we estimate the predicted class distribution based on  marginalizing the predictions over the classes.  For example, the outcome $\mathcal P=1$ is computed as:
\be
P(\mathcal P=1,M=m)&=&P(\mathcal P=1|\mathcal C=1,M=m)P(\mathcal C=1)\nonumber\\
&+& P(\mathcal P=1|\mathcal C=0,M=m)P(\mathcal C=0)\;,
\ee
 where $P(\mathcal C=1,M=m)=P(\mathcal C=1)$, and $P(\mathcal C=0,M=m)=P(\mathcal C=0)$, and we estimate the conditional probabilities for a given detector based on the detector's energy cutoff, since its predictions are based on the energy threshold.  

The conditional entropy is defined as:
\begin{equation}
 H(\mathcal P|\mathcal C,M=m)=\sum_p  P(p) \sum_c  P(p|c,M=m)\log{ P(p|c,M=m)},     
\end{equation}         
where the conditional probability $P(p|c,M=m)$ is a function of the $TPR$ and $FPR$ that is determined by the conditional detector, for example if $M=DC$ we have:
 \be\label{dcarray}
    P(1|1)=P(\mathcal P=1|\mathcal C=1, M=DC) &=  TPR \\ \nonumber
   P(1|0)=P(\mathcal P=1|\mathcal C=0, M=DC) &= FPR\\ \nonumber
   	P(0|1)=P(\mathcal P=0|\mathcal C=1, M=DC) &= 1-TPR \\ \nonumber
   P(0|0)=P(\mathcal P=0|\mathcal C=0, M=DC) &= 1-FPR \;.  \nonumber
  \ee

The TPR and FPR are a function of the conditional detector.  For the DC detector we defined the positives as the known 'proximal' sites, previously denoted as $D_{DC}$, and the negatives are the known 'distal' sites, $D_{DU}$. 

Two equations in \Eref{dcarray} are based on the FPR, and defined as: $P(\mathcal P=0|\mathcal C=0,DC)$ is the fraction of known 'distal' binding sites whose energy is above the energy threshold (true negatives divided by negatives),  and $P(\mathcal P=1|\mathcal C=0,DC)$ is the fraction of known 'distal' binding sites whose energy is below the energy threshold (false positives divided by negatives).  Two equations are based on the TPR, $P(\mathcal P=1|\mathcal C=1,DC)$ is the fraction of known 'proximal' binding sites whose energy is below the energy threshold, and of course $P(\mathcal P=0|\mathcal C=1,DC)=1-P(\mathcal P=1|\mathcal C=1,DC)$, which is the fraction of known `proximal' binding sites whose energy is above the energy threshold (false negatives divided by positives). 

For the DU detector, the DU positives are what DC would call a negative, hence DU's positive data are the known `distal' sites, and its negatives are the known `proximal' sites.  This is because DU should be firing when Twist is distal.  For example, the TPR now determines $P(\mathcal P=0|\mathcal C=0,DU)$, which is defined as the fraction of known `distal' binding sites whose DU's energy score is below the energy threshold (true positives divided by positives). The FPR for DU is now given by $P(\mathcal P=0|\mathcal C=1,DU)$, which is the fraction of known `proximal' binding sites whose energy is below the energy threshold (false positives divided by negatives).

\section*{Additional Experiment Supplement, Rerunning Model on CACATG Twist Motif}

We repeated our analysis of the known Dorsal sites with a different Twist motif (5'-CACATG).  We used the same sliding window as in the main text of 30 base pair shifts starting at [0,30]bp, and then incrementing to [31,60]bp etc.  We also used the same spacer window as in the main text ([0,30]bp) for the DC detector for making predictions of known Dorsal loci in the CRMs.

\begin{figure}[ht]
\includegraphics[width=6in]{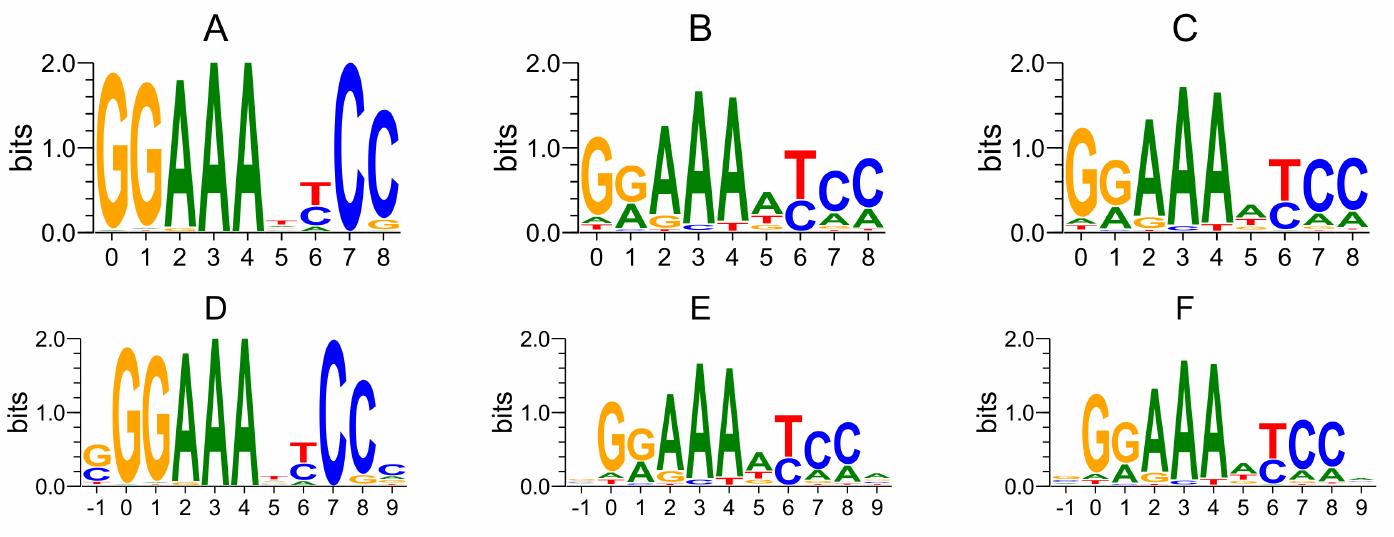}  
\caption{Logos generated for known Dorsal sites tested for adjacency to 5'-CACATG used as 'cooperative' class (DC) if in the [0,30]bp distance.  Logo A corresponds to the 'Dorsal Cooperative' class, it's total information content we calculated at 13.4bits.  Logo D is the exact same logo as A but we've appended one base-pair of flanking sequence onto the start and end of the site (hence, this logo starts at position -1).  Position 9 of this logo shows about a couple decibits of information relative to the background sequence and the position -1 contains a half bit of information.  Logo B is the 'Dorsal Uncooperative' class for the [0,30]bp window, which we calculated to have 9.4 bits information relative to the background (uniform distribution of bases), and logo E has added the flanking sites to the 'Dorsal Uncooperative' class.  Logo C is the CB motif with 9.7 bits of information relative to the background, which looks similar to the 'Dorsal Uncooperative' class at position 6 due to there being many more sites that prefer A to a T at this position amongst all the Dorsal sites in the network.  Logo F is the CB motif with the flanking sequence appended.}
\end{figure}

\par

  \begin{tabular}[b]{|c|c|c|c|}
\hline
spacer   &  [0,30]bp & (31,60]bp &(61,90]bp  \\ \hline
\multirow{2}{*}{}  Mutual Information \Eref{mi}&  0.38 & 0.28 & 0.04  \\
 Logodds ratio test -log(p value)& 4.8 & 0.17 & 0.66 \\
\hline
\end{tabular}\label{tablesuple}

\par

This  table's second row corresponds to the log odds ratio test based on CB's specificity set at $10^{-4}$, which corresponds to a CB energy of 2.5, as in the main text.  The DC detector did show better performance for this spacer window and corresponding energy cutoff.  We additionally ananlyzed the mutual information for the case that 5'-CACATGT was used as a motif for Twist, which would correspond to a subset of the sites found for 5'-CACATG.  For the 5'-CACATGT motif we found the mutual information was 0.3, 0.17, 0.0 for the three possible cases of the sliding window.

\newpage

\subsection{ROC curve}
 The ROC curve for the OR gate and the CB detector for 5'-CACATG Twist motif are displayed in Figure 6. 
 The detectors behave in a similar manner to the results presented in the main paper for the 5'-CAYATG motif.
\begin{figure}[!htbp]
\includegraphics[width=5in]{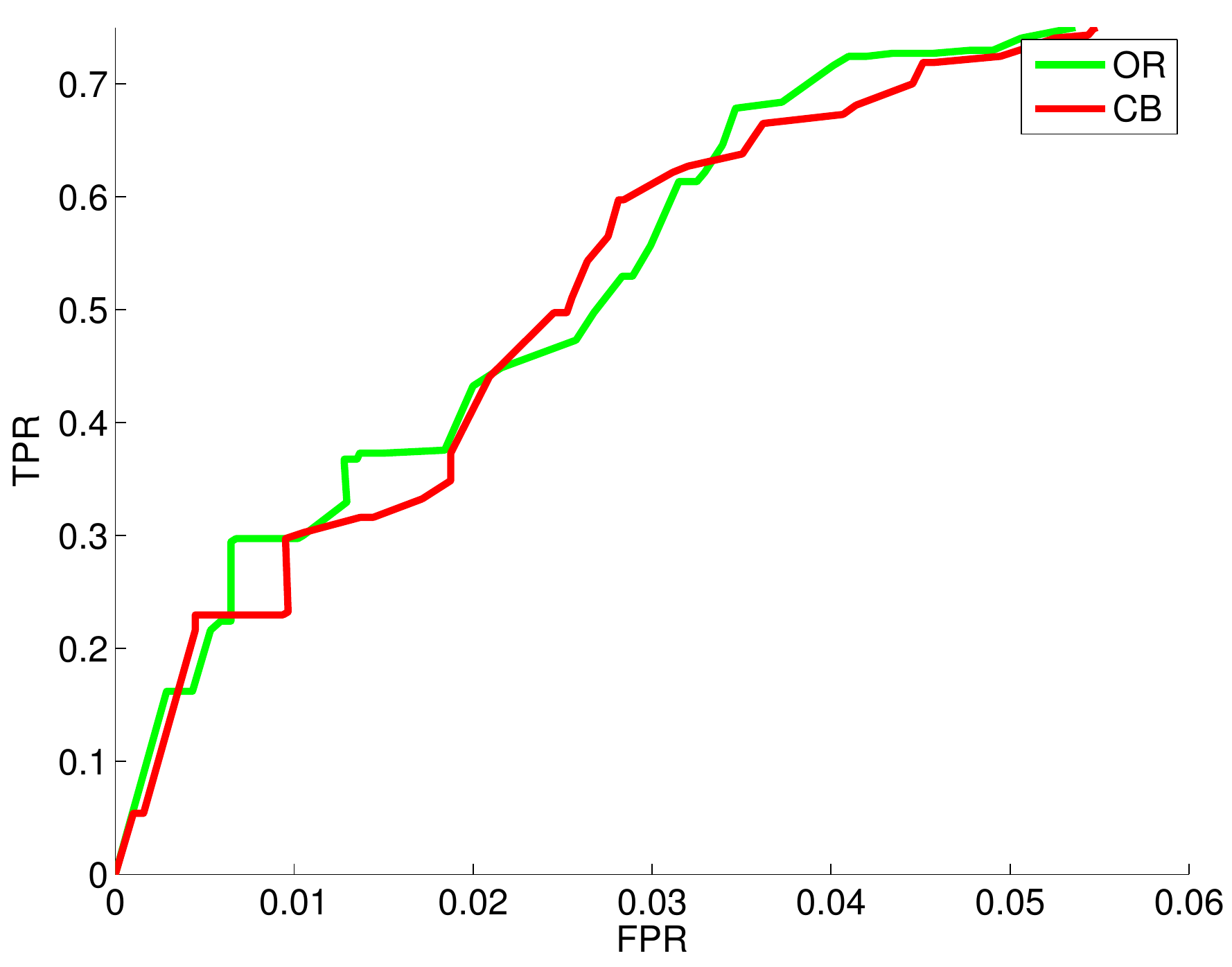}
\caption{ROC curves display the False positive rate (FPR) vs. True Positive Rate (TPR).  }
\end{figure}\label{rocfig2sup}

\newpage

\subsection{ Mutual Information between loci classes $C$ and detector predictions of classes $P$}
The mutual information $I(C;\mathcal P)$ for both conditional detectors based on a 5'-CACATG Twist motif in Figure 7
 shows similar behavior as the results in the main paper for the 5'-CAYATG Twist motif.   
\begin{figure}[!htbp]
\includegraphics[width=5in]{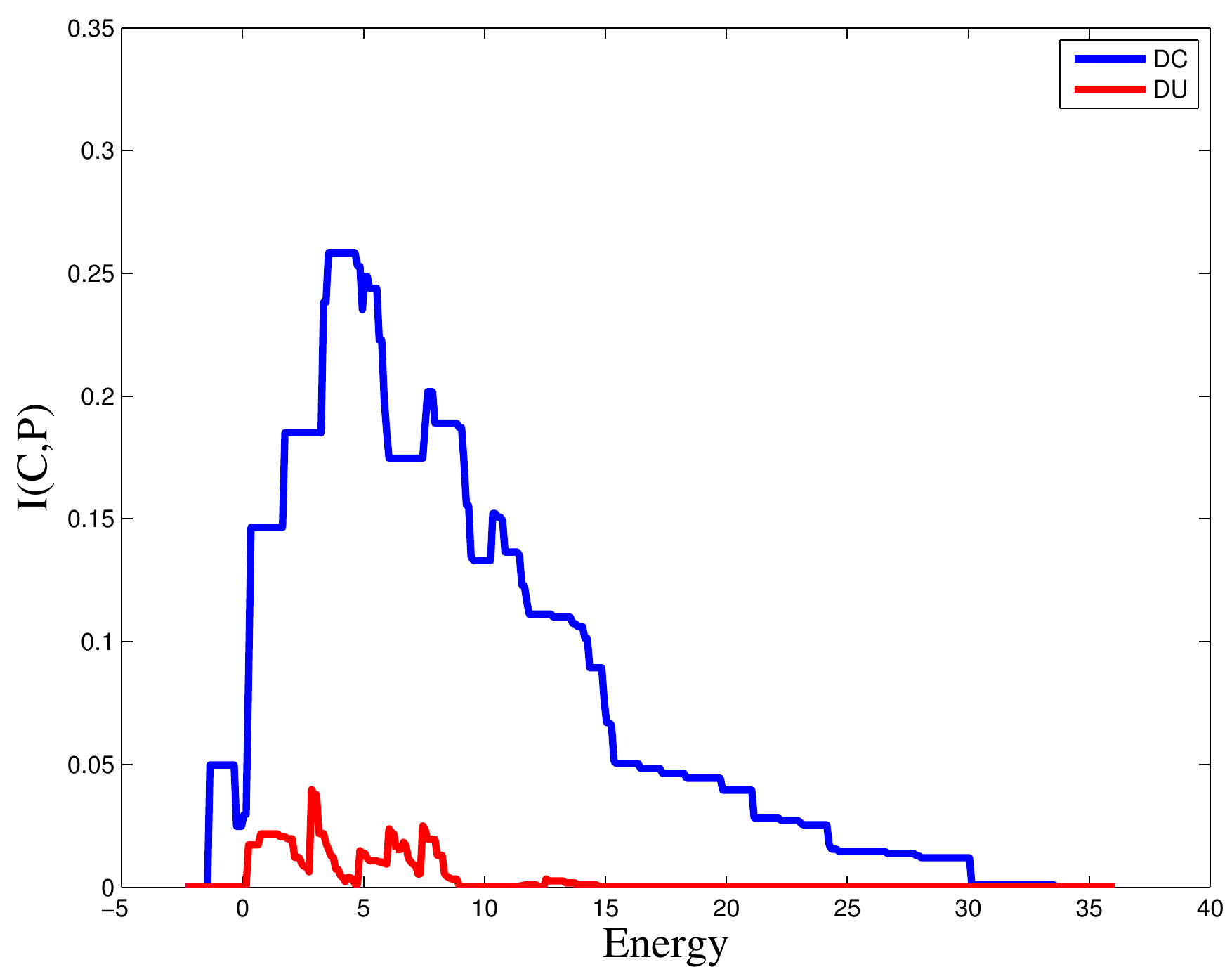}
\caption{The mutual information $I(C;\mathcal P)$ between the conditional detector's prediction's of class types (distal or proximal) and the known class types, as a function of the detection energy threshold is varied.  DC shows about 0.3 bits of information at about an energy cutoff of 4.  DU's score suggests that it does not do much better than random guessing in predicting class types.  }
\end{figure}\label{miCPsup}

\newpage
\subsection{Permutation test using rank sum statistic } 

The median energy of DC was 0.4, and the median energy of DU was 2.9.  The plot of the rank sum sampling distribution generated from random partitions of these data sets of size 66 and 356 respectively is below.
\begin{figure}[!htbp]
 \includegraphics[width=5in]{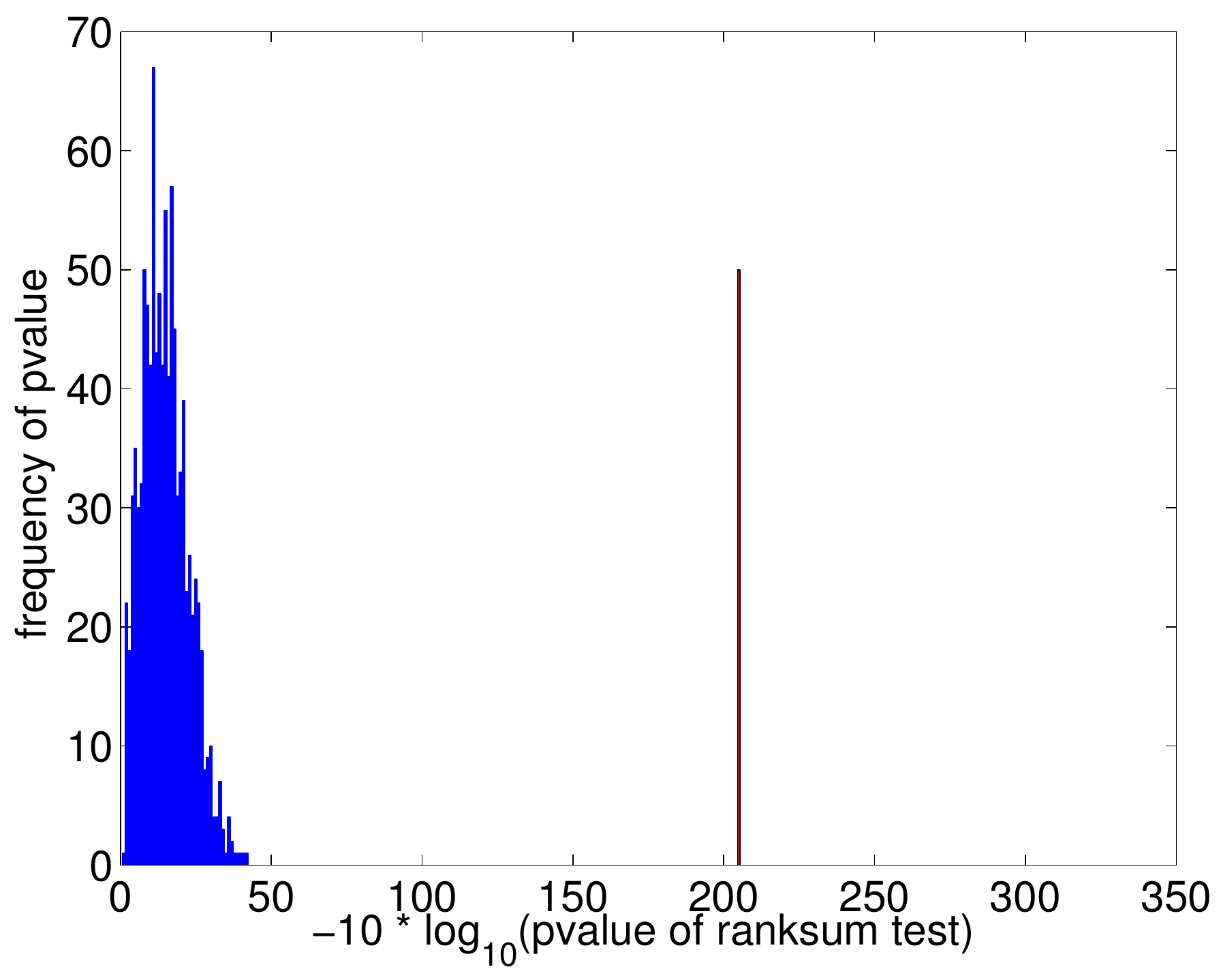}\\
  \caption{ Histogram of p-values of random partitions of the combined data set $\mathcal D_{\textbf{CB}}$, where the histogram bins were in units of 10*log(p value).  The p-value of the rank sum test for DC and DU median energies was about 205 in the scaled log units, which is the bar at the far left of the sampling distribution. }
\end{figure}\label{bb}
\newpage
\subsection{Entropy Bias}\label{entro}

The Bayesian estimate of the probability of $B$ with a Dirichlet prior using symmetric hyperparameter $\beta$  is $P(B)=\frac{n_{B}+\beta}{N+4*\beta}$\cite{MEP}.  To minimize the bias in the entropy we selected the value of $\beta$ that gave the smallest bias for the small sample regime.  Of course, we do not know the true entropy a-priori of the binding site distribution.  However, for the length 9 bp CB distribution of Dorsal binding site sequences we could initially estimate a ``known" PWM that could then be used to randomly generate synthetic binding sites.  Hence, using our best estimate of the CB PWM, we used the PWM to generate data sets of $N$ binding sites.  For each synthetic data we estimated a PWM with a predefined value of $\beta$, and thereby estimated the entropy as a function of $N$.  For each value of $N$, we generated $n$ replicate data sets, building a PWM for each set, thereby obtaining $n$ estimates of the entropy for each value of $N$.  We estimated the entropy based on a data sets of size $N$=[1,50], and $n=20$.  By repeating this experiment for values of $\beta$ in the domain $[10^{-5},0.1,0.2,0.25,0.5, 1]$ we found an empirical value of $\beta$ that best estimated the known entropy and energy.  We similarly repeated this for energy estimates, and found the least biased vale of $\beta=0.1$ for entropy and energy estimates.
\begin{figure}[!htbp]
\label{entropybias}
\includegraphics[width=5in]{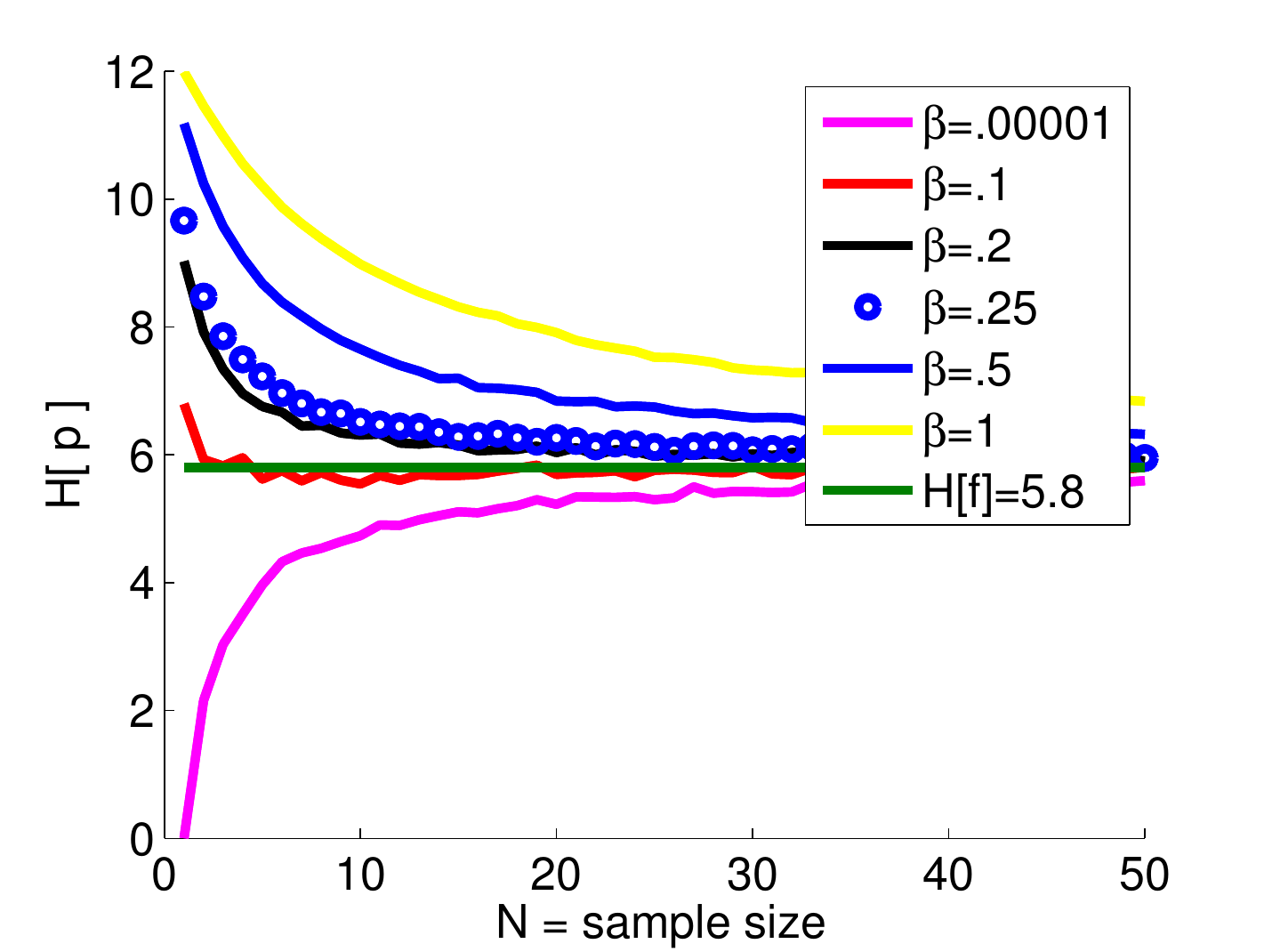}
\caption{ The probability distribution $p$ was estimated from $N$ random deviates of a known length 9 PWM that was built from $D_\textbf{CB}$ data.  The entropy of $p$, $H[p]=-\prod_{i=1}^9 \sum_B p(i,B) \log p(i,B)$ , where $i$ runs over the nine positions of the aligned $N$ sequence deviates, and $B$ runs over the bases, was computed for twenty replicates for each value of $N$ and plotted the average entropy over the twenty replicates as a function of $N$.  We computed the functional $H[p]$ as a function of $N$ for values of $\beta$ in the domain $[10^{-5},0.1,0.2,0.25,0.5, 1]$.  The known CB PWM had an entropy of 5.6 bits as shown by the green horizontal line, and we found an empirical value of $\beta$ that best estimated this 'known' entropy to be $\beta=0.1$ as shown by the red plot of the functional $H$ as a function of $N$.  We similarly repeated this for the functional energy estimates, and found the least biased vale of $\beta=0.1$ for entropy and energy estimates.}
\end{figure}

\end{document}